\documentclass[useAMS,usenatbib]{mn2e}%

\usepackage{amsmath}
\usepackage{graphicx}
\usepackage{lscape}
\usepackage{multirow}
\usepackage{deluxetable}

\usepackage{afterpage}

\def\rrg{${\it r}_{\rm g}$}
\def\rrin{${\it r}_{\rm in}$}

\def\laor{\rm{\sc laor}}
\def\diskbb{\rm {\sc diskbb}}

\def\refbhb{\rm{\sc refbhb}}

\def\reflionx{\rm{\sc reflionX}}
\def\nnh{${\it N}_{\rm H}$}
\def\ka{$K\alpha$}

\def\pcmsq{\hbox{$\rm\thinspace cm^{-2}$}}

\def\ergpcmsqps{\hbox{$\rm\thinspace erg~cm^{-2}~s^{-1}$}}

\def\ergcmps{\hbox{$\rm\thinspace erg~cm~s^{-1}$}}

\def\chisq{{\chi^{2}}}

\def\xspec{\hbox{\small XSPEC~\/}}

\def\j{\hbox{\rm J1550}}

\newcommand{\simpl}{{\sc simpl~}}

\newcommand{\powerlaw}{{\sc powerlaw~}}

\newcommand{\kerrbb}{{\sc kerrbb~}}
\newcommand{\kerrbbtwo}{{\sc kerrbb2~}}
\newcommand{\bhspec}{{\sc bhspec~}}

\newcommand{\tbabs}{{\sc tbabs}}
\newcommand{\phabs}{{\sc phabs}}
\newcommand{\simplr}{{\sc simpl-R}}
\newcommand{\simplC}{{\sc simplC}}
\newcommand{\kerrconv}{{\sc kerrconv}}
\newcommand{\ireflect}{{\sc ireflect}}
\newcommand{\reflionX}{{\sc reflionX}}
\newcommand{\smedge}{{\sc smedge}}
\newcommand{\crabcor}{{\sc crabcor}}
\newcommand{\gauss}{{\sc gauss}}

\newcommand{\spin}{a_{*}}
\newcommand{\lesssim}{\la}
\newcommand{\gtrsim}{\ga}

\newcommand{\simplb}{{\sc simpl}}

\newcommand{\kerrbbtwob}{{\sc kerrbb2}}

\newcommand{\Mdot}{\dot{M}}

\newcommand{\msun}{\rm M_{\sun}}

\newcommand{\rchinu}{\chi^{2}/\nu}
\newcommand{\fsc}{f_{\rm SC}}
\newcommand{\nh}{N_{\rm H}}

\newcommand{\kpc}{\rm kpc}
\newcommand{\kev}{\rm keV}
\newcommand{\keV}{\rm keV}

\newcommand{\ks}{\rm ks}

\newcommand{\nin}{n_{\rm in}}
\newcommand{\nout}{n_{\rm out}}

\newcommand{\rin}{r_{\rm in}}
\newcommand{\Rin}{R_{\rm in}}
\newcommand{\risco}{R_{\rm ISCO}}

\newcommand{\ledd}{L_{\rm Edd}}
\newcommand{\lledd}{L_{\rm D}/L_{\rm Edd}}
\newcommand{\lum}{L_{\rm D}}

\newcommand{\rxte}{{\it RXTE~}}
\newcommand{\chandra}{{\it Chandra~}}

\newcommand{\asca}{{\it ASCA~}}

\newcommand{\rxteb}{{\it RXTE}}

\newcommand{\gold}{{\em gold}}
\newcommand{\silver}{{\em silver}}

\title[The Spin of the Black Hole in XTE~J1550--564 ]{The Spin of the Black Hole Microquasar XTE J1550--564 via the Continuum-Fitting and Fe-Line Methods}

\author[Steiner et al.]{James F.\ Steiner$^{1}$\thanks{E-mail:
jsteiner@cfa.harvard.edu},
  Rubens C.\ Reis$^{2}$\thanks{E-mail:
rcr36@ast.cam.ac.uk},
  Jeffrey E.\ McClintock$^{1}$, 
  Ramesh Narayan$^{1}$,
\newauthor  Ronald A.\ Remillard$^{3}$, 
  Jerome A.\ Orosz$^{4}$, 
  Lijun Gou$^{1}$, 
  Andrew C.\ Fabian$^{2}$,
\newauthor  Manuel A.\ P.\ Torres$^{1,5}$\\
$^{1}${Harvard-Smithsonian Center for Astrophysics, 60
  Garden Street, Cambridge, MA 02138.}\\ 
$^{2}${Institute of
  Astronomy, Cambridge University, Madingley Road, Cambridge, CB3 0HA}\\
$^{3}${MIT Kavli Institute for Astrophysics and Space
  Research, MIT, 70 Vassar Street, Cambridge, MA 02139.}\\
$^{4}${Department of Astronomy, San Diego State University,
  5500 Campanile Drive, San Diego, CA 92182-1221.}\\
$^{5}${SRON, Netherlands Institute for Space Research,
    Sorbonnelaan 2, 3584 CA, Utrecht, The Netherlands.}}
\begin{document}

%\date{Accepted 1988 December 15. Received 1988 December 14; in original form 1988 October 11}
%\pagerange{\pageref{firstpage}--\pageref{lastpage}} \pubyear{2002}

\maketitle

%\label{firstpage}

%\email{jsteiner@cfa.harvard.edu}

\begin{abstract}

  We measure the spin of XTE J1550--564 in two ways: by modelling the
  thermal continuum spectrum of the accretion disc, and independently
  by modeling the broad red wing of the reflection fluorescence
  Fe-\ka\ line.  We find that the spin measurements conducted
  independently using both leading methods are in agreement with one
  another.  For the continuum-fitting analysis, we use a data sample
  consisting of several dozen \rxte spectra, and for the Fe-\ka\
  analysis, we use a pair of \asca spectra from a single epoch.  Our
  spin estimate for the black hole primary using the continuum-fitting
  method is $-0.11 < \spin < 0.71$ (90 per~cent confidence), with a
  most likely spin of $\spin = 0.34$.  In obtaining this result, we
  have thoroughly explored the dependence of the spin value on a wide
  range of model-dependent systematic errors and observational errors;
  our precision is limited by uncertainties in the distance and
  orbital inclination of the system.  For the Fe-line method, our
  estimate of spin is $\spin = 0.55^{+0.15}_{-0.22}$.  Combining these
  results, we conclude that the spin of this black hole is moderate,
  $\spin = 0.49 ^{+0.13}_{-0.20}$, which suggests that the jet of this
  microquasar is powered largely by its accretion disc rather than by
  the spin energy of the black hole.

\end{abstract}


\section{Introduction}\label{section:Intro}

During its principal 1998--1999 outburst cycle, the bright X-ray
transient XTE J1550--564 produced one of the most remarkable flare
events ever observed for a black hole binary.  For $\approx1$~day, the
source intensity rose fourfold relative to neighbouring plateau values,
reaching 6.8~Crab.  The flux in the dominant power-law component rose
by the same factor, and then just as quickly its intensity declined
\citep{Sobczak_2000, JEM_H1743}.  Four days later, AU-scale
superluminal radio jets were observed \citep{Hannikainen_2009}.  Their
separation angle ($\sim 255$ mas) and relative velocity ($\sim 65$ mas
d$^{-1}$) links the birth of these jets to the impulsive X-ray flare.
The subsequent detection of large-scale radio jets in 2000 led to the
discovery of relativistic X-ray jets \citep{Corbel_2002}.  All of the
available evidence strongly indicates that these pc-scale X-ray and
radio jets were produced by the unique 7-Crab flare event, and we
adopt this view.

The microquasar XTE J1550--564 (hereafter J1550) is further
distinguished by a pair of high-frequency X-ray oscillations with a
2:3 frequency ratio (184 and 276 Hz; \citealt{Remillard_2002}).
During its 1998--1999 eruption, J1550 displayed all of the active
accretion states: hard, steep power law (SPL), thermal dominant (TD)
and intermediate (INT; \citealt{RM06}).  The X-ray spectral and timing
properties of this source have been comprehensively studied by many
authors (e.g., \citealt{Sobczak_2000, Homan_2001, Remillard_2002,
  Kubota_Done_2004, Dunn_2010}), as have the properties of its radio
counterpart \citep{Corbel_2001, Xue_2008, Hannikainen_2009}.

Likewise, the optical counterpart of J1550 was the subject of a
comprehensive dynamical study by \citet{Orosz_2002}.  The measurement by
these authors of a large mass function immediately established J1550 as
a dynamically-confirmed black hole binary with a $\approx 10~M_{\odot}$
black hole primary in a 1.55-day orbit with a late G or early K
companion.  This dynamical model was recently revisited using new
photometric and spectroscopic data \citep{Orosz_Steiner_2010}.  Our
higher-resolution spectra (60 km~s$^{-1}$) revealed that the mass ratio
is extreme ($Q \approx 30$) and yielded a refined value of the mass
function, $f(M) = 7.65\pm0.38~M_{\odot}$.  Of central importance to the
present paper, \citet{Orosz_Steiner_2010} report accurate values of the
three key quantities that are essential for determining the spin of the
black hole via the continuum fitting method, namely the distance
$D=4.38_{-0.41}^{+0.58}$~kpc, black hole mass $M=9.10\pm0.61~M_{\odot}$,
and orbital inclination angle $i = 74\fdg7\pm3\fdg8$.

Currently, the two principal methods for measuring black hole
spin\footnote{Black hole spin is commonly expressed in terms of the
dimensionless quantity $a_* \equiv a/M = cJ/GM^2$, where $M$ and $J$ are
respectively the black hole mass and angular momentum
\citep{Shapiro_Teukolsky}.  Its limiting value is $a_*=+1$ ($-1$) for a
maximal Kerr hole rotating in a prograde (retrograde) sense relative to
the accretion disc; $a_*=0$ corresponds to a non-spinning Schwarzschild
hole.}  are modeling the thermal spectrum of the accretion disc
\citep{Zhang} and modeling the profile of the Fe-\ka\ line
\citep{Fabian_1989, laor}.  For both methods, spin is measured by
estimating the inner radius of the accretion disc $\rin \equiv R_{\rm
in}/M$ in standard GR units ($G = c = 1$).  $R_{\rm in}$ is identified
with the radius of the innermost stable circular orbit ($\risco$) of the
gravitational potential and is related to spin via a monotonic mapping
between the dimensionless ISCO radius $\risco/M$ and the dimensionless
spin parameter $a_*$ \citep{Shapiro_Teukolsky}.  Strong support for
linking $R_{\rm in}$ to $\risco$ is provided by decades of empirical
evidence that $\rin$ is constant in disc-dominated states of black hole
binaries (e.g., \citealt{Tanaka_Lewin}), as shown most compellingly in
our recent study of the persistent source LMC X-3 \citep{Steiner_lmcx3}.
Theoretical support for identifying $R_{\rm in}$ with $\risco$ is
provided by MHD simulations of thin accretion discs
(\citealt{Reynolds_Fabian_2008, Shafee_2008, Penna_2010}; but see
\citealt{Noble_2009, Noble_2010}).  In short, the relationship for thin
accretion discs between $\rin$, $\risco$ and $a_*$ is the foundation of
both the continuum-fitting and Fe-\ka\ methods of measuring spin.

In the continuum-fitting (CF) method, one determines $\risco$, and hence
$a_*$, via measurements of X-ray temperature and luminosity (i.e., using
X-ray flux, distance $D$ and inclination angle $i$) of the disc
emission.  In order to obtain reliable values of $a_*$, it is essential
to (1) select X-ray spectra that have a strong thermal component and (2)
have accurate estimates of $D$, $M$ and $i$, like those given above for
J1550.  In practice, we fit the X-ray spectrum of the black hole's
accretion disc to our version of the Novikov-Thorne thin accretion disc
model \citep{NT73, KERRBB, McClintock_2006} using an advanced treatment
of spectral hardening \citep{Davis_2005, BHSPEC}. In this way, we have
measured the spins of six other stellar black holes.  We find spins
ranging from $\spin\approx0.1$ \citep{Gou_2010} to $a_*>0.98$
(\citealt{McClintock_2006}); four other spin values are relatively high,
$\spin \approx0.7-0.9$ \citep{Shafee_2006, Liu_m33x7, Liu_2010, Gou_2009}.

%, while the value for J1550 is moderate, $a_*\approx0.5$.

In the Fe-\ka\ method, one determines $\risco$ by modeling the profile
of reflection-fluorescent features in the disc.  Most prominent is the
broad and skewed iron line, whose shape is determined by Doppler
effects, light bending, and gravitational redshift
\citep{Reynolds_Nowak_2003}.  Of central importance is the effect of the
redshift on the red wing of the line.  This wing extends to very low
energies for a rapidly rotating black hole ($a_* \sim 1$) because in
this case gas can orbit near the event horizon, deep in the potential
well of the black hole.  Relative to the CF method, measuring the extent
of this red wing in order to infer $a_*$ is hindered by the relative
faintness of the signal.  However, the Fe-\ka\ method has the virtues
that it is independent of $M$ and $D$, while the blue wing of the line
even allows an estimate of $i$.  What makes the Fe-\ka\ method
enormously important is that it is the primary approach to measuring the
spins of supermassive black holes in AGN.  The spins of several stellar
black holes \citep{Reis_2009, reis1752, Miller_2009, Blum_2009} and
supermassive black holes \citep{Brenneman_Reynolds, Schmoll_2009,
Miniutti_2009, fabian09nature, Zoghbi_2010} have been reported using the Fe line method
with values ranging from $\spin \approx 0$ to $\spin > 0.98$.

Knowledge of black hole spin has broad importance to astrophysics: For
example, spin is central to most of the many theories of relativistic
jets observed for both microquasars and AGN \citep{BZ77}, and it is
comparably important to collapsar models of long GRBs
\citep{Woosley_1993} and models of black hole formation and black hole
binary evolution \citep{Lee_Brown_Wijers_2002}.  Hierarchical models for
the growth of supermassive black holes require knowledge of the spin
distributions of the merging partners \citep{Volonteri_2005,
Berti_Volonteri_2008}, and the observed properties of AGN may be
strongly conditioned by black hole spin \citep{McNamara_2009,
Garofalo_2010, Tchekhovskoy_2010}.  Spin measurements are likewise
important to gravitational-wave astronomy in predicting the waveforms of
merging black holes \citep{Campanelli_2006}.  Knowledge of black hole
spin is becoming important to fundamental physics as well, and
enlivening questions are being asked: e.g., Is the No-Hair Theorem valid
and can it be tested \citep{Johannsen_2010}?  Do we live in a string
axiverse filled with light axions \citep{Arvanitaki_2010}?

There have been two prior estimates of the spin of J1550.  The first of
these, $a_*\approx0-0.1$, was obtained using the CF method and a sample
of ten \rxte\ spectra \citep{Davis_2006}.  This result was based on the
old dynamical model with approximate values of $M$, $i$, and $D$ (e.g.,
$D$ was uncertain by $\approx45$ per~cent; \citealt{Orosz_2002}).  We improve
upon the work of Davis et al.\ by using our new dynamical model (e.g.,
with its fourfold better determination of $D$) and a $\approx6$-times
larger sample of \rxte spectra, and by our detailed treatment of
observational and model-dependent uncertainties.  A second
`preliminary' spin estimate, $a_*\approx0.76$, was obtained using
a spectral model that self-consistently merged the disc-continuum
and Fe-\ka\ components, in which the spin result for \j\ was
predominately determined by the Fe-\ka\ model \citep{Miller_2009}.  We
will treat the data considered in their study, and additional data;
however, unlike this exploratory and preliminary study of eight
black-hole systems (including J1550), we fixedly focus on J1550 and so
are able to refine and improve upon their work.

In this Paper, we present the spin of \j\ on two fronts.  After
introducing the data sets (\S~\ref{section:Observations}), we begin by
first applying the CF technique
(\S\S~\ref{section:analysis},\ref{section:self-consistent}).  This
work is complemented with a thorough exploration of the statistical
and systematic uncertainties inherent to the CF method
(\S~\ref{section:systematics}).  Next, we measure the spin of J1550
using the Fe-\ka\ technique (\S~\ref{reflection}), and we finish with
a discussion of the results (\S~\ref{section:discussion}) and our
conclusions (\S~\ref{section:conclusion}).

%%%%%%%%%%%%%%%%%%%%%%%%%%%%%%%%%%%%%%%%%%%%%%%%%%%%%%%%%%%%%%%%%%%%%%%%%%

\section{Observations}\label{section:Observations}

\begin{figure}
{\includegraphics[clip=true, angle=90,width=8.85cm]{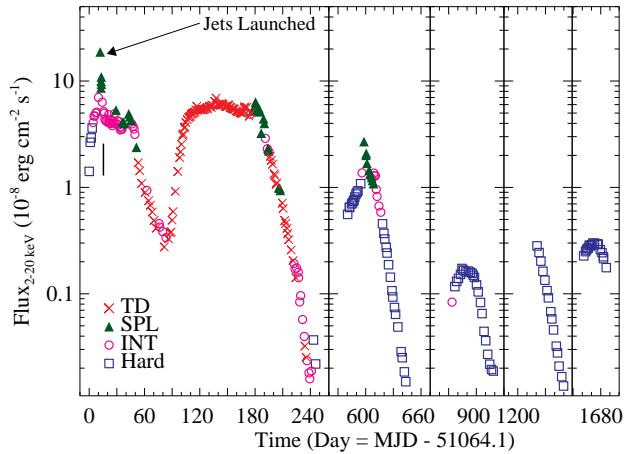}}
\caption{A spectral-state encoded 2--20 \keV\ light curve showing all
  five outburst cycles of J1550.  Most disc-dominated data were obtained
  during the primary outburst in 1998--1999.  The time of the \asca
  observation, which is analysed in \S~\ref{reflection}, is marked at
  day 15 by a solid black line.  The powerful 7-Crab flare near day 12
  was responsible for ejection of superluminal radio
  jets.}\label{fig:state-lc}
\end{figure}

The primary data set used in this study is a compendium of 347 {\it
Rossi X-ray Timing Explorer} (\rxteb) observations.  The data include
those obtained during the bright discovery outburst in 1998--1999 on
through four additional minor outbursts, the last ending in mid-2003.  A
light curve of the flux, showing the spectral evolution of the source,
is presented in Figure~\ref{fig:state-lc}, where the spectral state
assignments have been determined using precisely the model, procedures
and criteria described in \citet{RM06}.  Thermal dominant data, which
are of primary importance for CF spin measurements, were obtained
exclusively during the first outburst cycle.

\rxte spectral data are collected using exclusively PCU-2, the best
calibrated of the five Proportional Counter Array (PCA) detectors.
Spectra are individually obtained by grouping sequential observations
into approximately half-day bins with a typical exposure time $\sim
3~\ks$.  We follow the procedures described in
\citet{McClintock_2006}: The data are background subtracted, a
customary 1 per~cent systematic error in the data count rates is included,
and a dead time correction ranging from approximately 1 to 20 per~cent is
applied.  Spectra are fitted over the energy range 2.55--45~keV using
XSPEC version 12.4--12.6 (the software used throughout this Paper;
\citealt{XSPEC}).  Calibration over this range is achieved using the
latest version of PCARMF (v11.7)\footnote{
  http://www.universe.nasa.gov/xrays/programs/rxte/pca/doc/\\rmf/pcarmf-11.7/}.

A linear collimator correction (assuming an ideal $1\degr$ triangular
response; \citealt{Jahoda_2006}) has been applied to the data set to
account for a series of offsets in the PCA pointing.  Specifically, we
normalised the flux upward by 4.4 per~cent (a 2.\arcmin67 offset) during the
1998--1999 outburst, and 7.1 per~cent (4.\arcmin28 offset) during the 2000
outburst.  After April 2001 the correction is just $\sim 0.1$ per~cent.  In
addition to these global corrections, a handful of observations taken
between 1998 September 7--9 and on 1999 January 6 were off-target for
unknown reasons by $\approx 0.2\degr$ and required us to make large
corrections (for details, see \citealt{Steiner_2009}).  We estimate that
the uncertainty in these flux corrections is no more than 1--2 per~cent, 
which has a negligible impact on our spectral fitting results.

In addition to the \rxte data set, we also include a 25\ \ks\ \asca
Gas Imaging Spectrometer (GIS) observation taken on UT 1998 September
23 when the source was in an intermediate state (see the observation
time in Fig.~\ref{fig:state-lc} marked by the black vertical line).
Following \citet{Miller_2009} and \citet{Miller_2005_j1550}, we use
these \asca data, with twice the resolution of the \rxte PCA data, to
examine the iron line.  However, we do not report a CF analysis on
these data because the Compton component is too strong.  We use
standard data products for the GIS-2 and GIS-3 spectra with version
4.0 response matrices.  The two spectra are fitted jointly over the
1--10~\kev\ energy range, and in \S~\ref{reflection} we present our
analysis of these data using the Fe-\ka\ method.

%%%%%%%%%%%%%%%%%%%%%%%%%%%%%%%%%%%%%%%%%%%%%%%%%%%%%%%%%%%%%

\section{Continuum-Fitting Analysis}\label{section:analysis}

We first enumerate our CF data selection requirements and define two
tiers of data quality.  Next, we introduce the first and principal of
three Comptonised accretion-disc models which are applied to the \rxte
data set.  Then, in the following section we introduce two alternative
models that differ principally in their treatment of the Compton
reflection component.  We find very close agreement in the spin
estimate using all three models.

\subsection{ \rxte  Data Selection}\label{subsec:selection}

We first identify two tiers of quality in our data: first-class `gold'
spectra are selected from just the strongly disc-dominated TD-state
observations (the most conducive to CF modeling; e.g.,
\citealt{Shafee_2006}).  We additionally consider a broader set of
second-tier `silver' spectra which are chosen from either SPL or
intermediate states.  In order for a spectrum to be assigned to either
category, it must pass the Comptonisation strength, quality, and
luminosity filters described below.

In our previous work on \j\ and the black-hole candidate H1743--322
(\citealt{Steiner_2009}), we showed that one can analyse a regime of
SPL- and intermediate-state spectra to obtain values of $r_{\rm in}$ (or
equivalently $\spin$) that are consistent with those obtained in the
({\em gold}) TD state.  Based upon this work, we select soft thermal
spectra with a power-law normalisation $\fsc < 25$ per~cent, which at $\fsc
\approx 25$ per~cent corresponds to a power-law component roughly an order of
magnitude stronger than that observed in the TD state.  Data fulfilling
this broader definition (which are not already in the TD {\em gold}
class) are candidates for the \silver\ class.

To be admitted into either category, we require that data meet two
additional criteria pertaining to the fit quality and disc luminosity.
For quality screening, we adopt a critical goodness-of-fit, defined as
$\rchinu < 2$, and also require that the inner-disc radius measurement
have a precision better than $\rin/\sigma_{\rin} = 5$.  We further
require that the luminosity of the soft thermal component lie in the
range 5--30 per~cent $\ledd$.  The upper luminosity threshold is required by
our thin-disc model \citep{McClintock_2006, Penna_2010}.  The lower
threshold eliminates a regime in which the accretion flow properties
may be changing, due to the presence of either advective flows
(\citealt{Esin_1997}) or other modes of coronal feedback (e.g.,
\citealt{Beloborodov_1999}).

\subsection{Results I: Continuum Fitting using \smedge }\label{subsec:resultsI}

In selecting our data and for determining the spin, we employ a variant
of the principal model that we used in our earlier study of J1550
and H1743--322 (hereafter, Model~S; \citealt{Steiner_2009}): \crabcor$\times$\tbabs$\times$\smedge(\simplb$\otimes$\kerrbbtwob).

Here we use \tbabs\ \citep{Wilms_2000} in place of \phabs\ to describe
the low-energy absorption component.  We fix the column density to the
precise value determined using \chandra grating data, $\nh =
8.0^{+0.4}_{-0.3} \times 10^{21}\pcmsq$ (90 per~cent confidence;
\citealt{Miller_2003_j1550}).  The custom multiplicative component
\crabcor\ simply corrects the response of the PCA detector using our
standard reference spectrum of the Crab (see \citealt{Steiner_lmcx3,
  Toor_Seward}).

The key component of this model is \kerrbbtwo \citep{McClintock_2006}, a
fully relativistic thin accretion-disc model, which includes
self-irradiation of the disc (`returning radiation') and limb
darkening \citep{KERRBB}.  The effects of spectral hardening are
incorporated via a pair of look-up tables for the hardening factor $f$
\citep{Davis_2005,BHSPEC} corresponding to two representative values of
the viscosity parameter: $\alpha = 0.01$ and $\alpha = 0.1$.  Here and
throughout, motivated by the results of both observational data and
global GRMHD simulations (\citealt{Penna_2010, King_2007}, and
references therein), we adopt $\alpha = 0.1$ as our fiducial value.
Following our previous work, we adopt a zero-torque inner boundary
condition and assume alignment of the black hole spin axis with the
binary orbital plane; we turn on both limb-darkening and returning
radiation flags and fix the normalisation to one.  We fix the input
parameters $M$, $i$ and $ D$ to their best-fitting values
(\S~\ref{section:Intro}).  The model \kerrbbtwo has just two fit
parameters, namely the black hole spin parameter $a_*$ and the mass
accretion rate $\dot M$, or equivalently, $\rin$ and the
Eddington-scaled bolometric disc luminosity $\lledd$ \citep{McClintock_2006}.

\begin{figure}
\centering{\includegraphics[clip=true, angle=90,width=8.8cm]{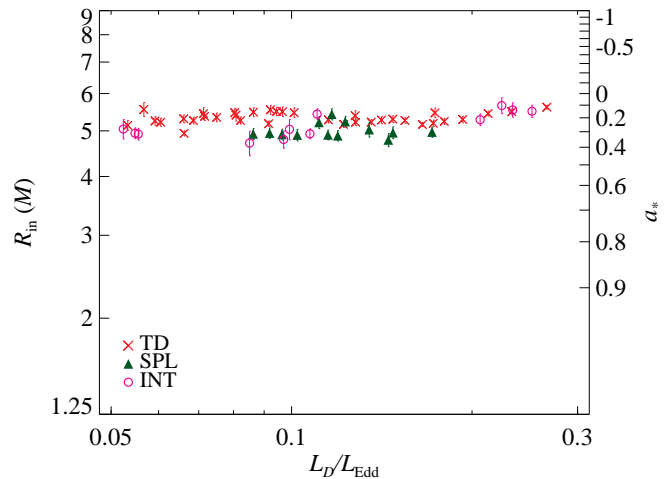}}
\caption{The spin, expressed both in terms of $\Rin$ and $\spin$,
versus luminosity.  The TD data comprise the \gold\ data set and
the intermediate and SPL data the \silver\ data set.  The mean
value of $\Rin$ is in agreement for the two data sets to within
$\approx 5$ per~cent even though the Compton component is much stronger for
the \silver\ data.}\label{fig:lumspin}
\end{figure}

We model the power-law component by convolving the thermal component
with \simpl (\citealt{Steiner_simpl}), a model that mimics the physics of
Compton scattering of thermal disc photons by a hot corona.  The model
\simpl converts a fraction $\fsc$ of the seed photons into a power law
with photon index $\Gamma$.  We use the standard, upscattering-only
version.  For the reflected component, we assume here that the disc
elastically backscatters all incident Compton photons (generated by
\simplb), apart from a broad iron absorption edge feature that is
modelled phenomenologically using \smedge\ \citep{SMEDGE}.  The
parameters of \smedge\ are the edge energy $E_{\rm Edge}$ (fitted from
7--9~\keV), its optical depth $\tau_{\rm max}$ (unconstrained in the
fit), and the width of the feature $W_{\rm Edge}$ (fixed at 7 keV).  In
the section that follows, we consider two models of reflection that are
more physically motivated.

Applying our selection criteria to the full spectral model yields 35
\gold\ spectra, where most of the winnowing is a result of our thin-disc
limit on the intrinsic luminosity (i.e., prior to scattering) of the
accretion-disc component: $\lledd<0.3$.  We additionally select 25
\silver\ spectra, 13 of which correspond to SPL-state observations and
12 to intermediate-state observations.  Our spectral-fitting results are
summarized in Table~\ref{tab:smedge} (\gold\ spectra correspond to
entries 1--35 and \silver\ to entries 36--60).

For all these selected data, in Figure~\ref{fig:lumspin} we plot
$\spin$ versus the luminosity of the disc component $\lledd$.  In a
departure from our earlier work, in addition to $\spin$, we also plot
the inner disc radius $\Rin$.  The two quantities are equivalent in
the sense that they are simply related to each other via a monotonic
analytical formula (\S~\ref{section:Intro}).  We have chosen to also
show $\Rin$ because it is the quantity that is more directly
determined via continuum fitting.

Figure~\ref{fig:lumspin} shows that the \gold\ and \silver\ data sets
give result that are in good agreement.  The net weighted result for
the combined data set is $\spin = 0.23 \pm 0.07$ ($\rin = 5.22 \pm
0.24$).  The \gold\ data give a slightly lower value for the spin,
$\spin = 0.20$, than do the \silver\ data, $\spin = 0.27$; the
corresponding shifts in the mean value of $\rin$ are respectively $+2$
per~cent and $-3$ per~cent.

Although the data are clustered within a few per cent of a central
value of $\rin$, there is a small $\sim 5$ per~cent increase with increasing
$\lum$, which is most pronounced for $\lledd>0.2$, a pattern that has
been previously observed for other sources (e.g., GRS~1915+105,
\citealt{McClintock_2006}; LMC X--3, \citealt{Steiner_lmcx3}).  We
tentatively attribute this effect to a thickening of the disc with
luminosity and the limitation of our razor-thin disc model.

  \begin{deluxetable}{llcccccccccccccccccc} 
  %\rotate 
  \tabletypesize{\scriptsize} 
  \tablecolumns{          20}
  \tablewidth{0pc}  
  \tablecaption{Model~S Continuum-Fitting Results}
  \tablehead{\colhead{N} & \colhead{MJD} & \colhead{$\frac{L_D}{L_{\tiny{\textrm{Edd}}}}$} &  \multicolumn{2}{c}{\smedge}  &  &  \multicolumn{2}{c}{\simpl}  &  &  \multicolumn{2}{c}{\kerrbbtwo} & \colhead{$ \chi ^2 _\nu $/DOF}  & \colhead{State} \\ \cline{4-5} \cline{7-8} \cline{10-11}  \\  & & &  \colhead{$E_{\rm Edge} ({\rm keV})$}  &  \colhead{$\tau _{\rm max}$}  &  &  \colhead{$\Gamma$}  &  \colhead{$f_{sc}$}  &  &  \colhead{$\spin$}  &  \colhead{$\Mdot (10^{18} {\rm g/s})$} } 
 \startdata  \label{tab:smedge} 

1&51117.4&0.17&$      8.32\pm      0.14$&$      1.11\pm      0.16$&&$      2.06\pm      0.03$&$     0.029\pm     0.002$&&$     0.241\pm     0.023$&$      3.30\pm      0.11$&0.7/74&TD&\\
2&51119.0&0.15&$      8.12\pm      0.11$&$      1.34\pm      0.16$&&$      2.11\pm      0.03$&$     0.030\pm     0.002$&&$     0.220\pm     0.021$&$      3.01\pm      0.09$&0.8/74&TD&\\
3&51121.0&0.14&$      8.26\pm      0.15$&$      1.45\pm      0.25$&&$      2.15\pm      0.06$&$     0.011\pm     0.001$&&$     0.232\pm     0.019$&$      2.62\pm      0.07$&0.7/74&TD&\\
4&51124.7&0.12&$      8.08\pm      0.13$&$      1.72\pm      0.18$&&$      2.15\pm      0.04$&$     0.023\pm     0.001$&&$     0.215\pm     0.023$&$      2.25\pm      0.08$&0.6/74&TD&\\
5&51128.6&0.10&$      7.86\pm      0.10$&$      1.71\pm      0.12$&&$      2.17\pm      0.03$&$     0.036\pm     0.001$&&$     0.161\pm     0.039$&$      2.04\pm      0.10$&1.0/74&TD&\\
6&51130.5&0.09&$      7.98\pm      0.09$&$      1.71\pm      0.09$&&$      2.17\pm      0.02$&$     0.038\pm     0.001$&&$     0.151\pm     0.034$&$      1.92\pm      0.09$&1.3/74&TD&\\
7&51132.5&0.09&$      7.84\pm      0.10$&$      1.72\pm      0.11$&&$      2.16\pm      0.02$&$     0.035\pm     0.001$&&$     0.157\pm     0.034$&$      1.75\pm      0.09$&0.8/74&TD&\\
8&51134.5&0.08&$      7.74\pm      0.10$&$      1.85\pm      0.12$&&$      2.28\pm      0.03$&$     0.027\pm     0.001$&&$     0.161\pm     0.035$&$      1.63\pm      0.08$&0.8/74&TD&\\
9&51136.9&0.07&$      7.84\pm      0.09$&$      2.02\pm      0.11$&&$      2.16\pm      0.03$&$     0.025\pm     0.001$&&$     0.189\pm     0.031$&$      1.42\pm      0.07$&0.9/74&TD&\\
10&51145.5&0.05&$      7.92\pm      0.10$&$      1.80\pm      0.11$&&$      2.05\pm      0.02$&$     0.025\pm     0.001$&&$     0.255\pm     0.035$&$      1.01\pm      0.06$&0.7/74&TD&\\
11&51150.1&0.06&$      7.60\pm      0.12$&$      1.80\pm      0.14$&&$      2.08\pm      0.03$&$     0.030\pm     0.001$&&$     0.223\pm     0.031$&$      1.15\pm      0.06$&0.9/74&TD&\\
12&51152.1&0.07&$      7.94\pm      0.14$&$      2.61\pm      0.26$&&$      2.13\pm      0.07$&$     0.015\pm     0.001$&&$     0.220\pm     0.024$&$      1.33\pm      0.05$&0.8/74&TD&\\
13&51152.9&0.07&$      8.19\pm      0.11$&$      3.35\pm      0.28$&&$      2.20\pm      0.08$&$     0.010\pm     0.001$&&$     0.311\pm     0.017$&$      1.21\pm      0.04$&0.8/74&TD&\\
14&51154.0&0.08&$      7.84\pm      0.20$&$      2.78\pm      0.45$&&$      2.39\pm      0.13$&$     0.012\pm     0.002$&&$     0.220\pm     0.026$&$      1.60\pm      0.06$&0.7/74&TD&\\
15&51155.1&0.09&$      8.44\pm      0.22$&$      2.07\pm      0.39$&&$      2.43\pm      0.13$&$     0.008\pm     0.001$&&$     0.244\pm     0.021$&$      1.75\pm      0.06$&0.7/74&TD&\\
16&51157.6&0.12&$      8.59\pm      0.41$&$      1.32\pm      0.57$&&$      2.75\pm      0.25$&$     0.007\pm     0.002$&&$     0.249\pm     0.022$&$      2.33\pm      0.07$&0.8/74&TD&\\
17&51160.3&0.17&$      7.00\pm      2.96$&$      0.14\pm      0.22$&&$      1.94\pm      0.27$&$     0.002\pm     0.001$&&$     0.249\pm     0.015$&$      3.15\pm      0.07$&0.7/74&TD&\\
18&51162.2&0.21&$      7.59\pm      0.14$&$      1.38\pm      0.26$&&$      3.11\pm      0.65$&$     0.004\pm     0.004$&&$     0.167\pm     0.020$&$      4.29\pm      0.10$&1.1/74&TD&\\
19&51163.2&0.23&$      7.21\pm      0.16$&$      1.58\pm      0.33$&&$      3.20\pm      1.34$&$     0.004\pm     0.008$&&$     0.153\pm     0.023$&$      4.74\pm      0.13$&0.7/74&TD&\\
20&51164.2&0.27&$      7.00\pm      0.32$&$      1.34\pm      0.21$&&$      2.33\pm      1.12$&$     0.001\pm     0.003$&&$     0.117\pm     0.016$&$      5.55\pm      0.11$&1.0/74&TD&\\
21&51260.6&0.19&$      8.12\pm      0.12$&$      1.12\pm      0.20$&&$      2.06\pm      0.04$&$     0.030\pm     0.002$&&$     0.213\pm     0.025$&$      3.77\pm      0.12$&0.4/66&TD&\\
22&51261.8&0.18&$      8.39\pm      0.16$&$      1.00\pm      0.18$&&$      2.03\pm      0.03$&$     0.029\pm     0.002$&&$     0.227\pm     0.023$&$      3.48\pm      0.11$&0.7/66&TD&\\
23&51263.1&0.17&$      8.12\pm      0.12$&$      1.42\pm      0.16$&&$      2.13\pm      0.03$&$     0.036\pm     0.002$&&$     0.162\pm     0.038$&$      3.51\pm      0.16$&0.6/66&TD&\\
24&51264.8&0.15&$      8.21\pm      0.10$&$      1.55\pm      0.15$&&$      2.07\pm      0.03$&$     0.031\pm     0.001$&&$     0.210\pm     0.025$&$      2.89\pm      0.10$&0.7/66&TD&\\
25&51265.6&0.14&$      8.28\pm      0.12$&$      1.40\pm      0.14$&&$      2.09\pm      0.03$&$     0.035\pm     0.002$&&$     0.217\pm     0.025$&$      2.75\pm      0.10$&0.8/66&TD&\\
26&51266.9&0.13&$      8.18\pm      0.12$&$      1.60\pm      0.17$&&$      2.12\pm      0.03$&$     0.024\pm     0.001$&&$     0.233\pm     0.024$&$      2.46\pm      0.09$&0.5/66&TD&\\
27&51267.6&0.13&$      8.14\pm      0.09$&$      1.80\pm      0.11$&&$      2.13\pm      0.02$&$     0.040\pm     0.001$&&$     0.184\pm     0.038$&$      2.54\pm      0.12$&0.9/66&TD&\\
28&51273.6&0.10&$      8.10\pm      0.08$&$      1.82\pm      0.09$&&$      2.17\pm      0.02$&$     0.040\pm     0.001$&&$     0.153\pm     0.038$&$      1.96\pm      0.10$&0.7/66&TD&\\
29&51274.5&0.09&$      7.88\pm      0.10$&$      1.91\pm      0.12$&&$      2.12\pm      0.03$&$     0.033\pm     0.001$&&$     0.138\pm     0.036$&$      1.89\pm      0.10$&0.7/66&TD&\\
30&51276.3&0.08&$      7.96\pm      0.12$&$      2.64\pm      0.22$&&$      2.04\pm      0.05$&$     0.019\pm     0.001$&&$     0.179\pm     0.047$&$      1.61\pm      0.10$&0.6/66&TD&\\
31&51277.4&0.07&$      7.98\pm      0.10$&$      2.20\pm      0.14$&&$      2.07\pm      0.03$&$     0.026\pm     0.001$&&$     0.197\pm     0.033$&$      1.48\pm      0.08$&1.0/66&TD&\\
32&51278.7&0.07&$      7.82\pm      0.13$&$      2.03\pm      0.16$&&$      2.11\pm      0.04$&$     0.027\pm     0.001$&&$     0.168\pm     0.047$&$      1.44\pm      0.10$&0.7/66&TD&\\
33&51279.6&0.07&$      7.93\pm      0.10$&$      2.16\pm      0.13$&&$      2.06\pm      0.03$&$     0.023\pm     0.001$&&$     0.207\pm     0.029$&$      1.30\pm      0.06$&0.8/66&TD&\\
34&51280.6&0.06&$      8.11\pm      0.12$&$      2.19\pm      0.17$&&$      2.06\pm      0.04$&$     0.021\pm     0.001$&&$     0.233\pm     0.030$&$      1.17\pm      0.06$&0.7/66&TD&\\
35&51283.2&0.06&$      7.45\pm      0.20$&$      1.57\pm      0.22$&&$      2.21\pm      0.05$&$     0.027\pm     0.002$&&$     0.135\pm     0.057$&$      1.17\pm      0.10$&0.6/66&TD&\\
\hline
36&51110.3&0.25&$      8.32\pm      0.08$&$      1.11\pm      0.09$&&$      2.55\pm      0.02$&$     0.223\pm     0.007$&&$     0.149\pm     0.049$&$      5.13\pm      0.25$&0.8/74&INT&\\
37&51111.6&0.23&$      8.14\pm      0.10$&$      1.14\pm      0.10$&&$      2.50\pm      0.02$&$     0.240\pm     0.009$&&$     0.139\pm     0.058$&$      4.79\pm      0.27$&0.8/74&INT&\\
38&51112.8&0.22&$      8.14\pm      0.09$&$      1.16\pm      0.09$&&$      2.51\pm      0.02$&$     0.249\pm     0.008$&&$     0.103\pm     0.065$&$      4.70\pm      0.29$&0.8/74&INT&\\
39&51113.7&0.21&$      8.32\pm      0.09$&$      1.24\pm      0.10$&&$      2.49\pm      0.02$&$     0.208\pm     0.008$&&$     0.214\pm     0.044$&$      4.03\pm      0.20$&1.0/74&INT&\\
40&51115.3&0.17&$      8.36\pm      0.08$&$      1.33\pm      0.09$&&$      2.42\pm      0.02$&$     0.153\pm     0.005$&&$     0.303\pm     0.035$&$      3.14\pm      0.14$&0.8/74&SPL&\\
41&51126.6&0.11&$      8.19\pm      0.08$&$      1.67\pm      0.08$&&$      2.29\pm      0.02$&$     0.059\pm     0.001$&&$     0.172\pm     0.041$&$      2.21\pm      0.11$&1.0/74&INT&\\
42&51140.0&0.06&$      8.01\pm      0.09$&$      1.75\pm      0.10$&&$      2.17\pm      0.02$&$     0.086\pm     0.002$&&$     0.316\pm     0.039$&$      1.01\pm      0.07$&1.0/74&INT&\\
43&51140.7&0.05&$      8.13\pm      0.08$&$      1.77\pm      0.09$&&$      2.19\pm      0.02$&$     0.078\pm     0.002$&&$     0.310\pm     0.036$&$      1.00\pm      0.07$&1.1/74&INT&\\
44&51143.8&0.05&$      8.04\pm      0.09$&$      1.69\pm      0.10$&&$      2.21\pm      0.02$&$     0.081\pm     0.003$&&$     0.281\pm     0.068$&$      0.98\pm      0.10$&0.8/74&INT&\\
45&51269.7&0.12&$      8.25\pm      0.08$&$      1.83\pm      0.09$&&$      2.25\pm      0.02$&$     0.052\pm     0.001$&&$     0.173\pm     0.043$&$      2.34\pm      0.13$&1.0/66&SPL&\\
46&51270.8&0.09&$      8.30\pm      0.08$&$      1.64\pm      0.08$&&$      2.28\pm      0.01$&$     0.110\pm     0.002$&&$     0.309\pm     0.035$&$      1.68\pm      0.09$&1.0/66&SPL&\\
47&51271.4&0.09&$      8.34\pm      0.09$&$      1.48\pm      0.08$&&$      2.30\pm      0.02$&$     0.120\pm     0.003$&&$     0.315\pm     0.036$&$      1.57\pm      0.09$&0.9/66&SPL&\\
48&51664.4&0.15&$      8.19\pm      0.08$&$      1.50\pm      0.10$&&$      2.45\pm      0.02$&$     0.165\pm     0.005$&&$     0.307\pm     0.041$&$      2.70\pm      0.15$&0.8/66&SPL&\\
49&51664.7&0.15&$      8.39\pm      0.08$&$      1.42\pm      0.09$&&$      2.42\pm      0.02$&$     0.173\pm     0.005$&&$     0.354\pm     0.042$&$      2.56\pm      0.15$&0.9/66&SPL&\\
50&51665.4&0.13&$      8.11\pm      0.09$&$      1.46\pm      0.09$&&$      2.42\pm      0.02$&$     0.145\pm     0.004$&&$     0.286\pm     0.051$&$      2.50\pm      0.16$&0.8/66&SPL&\\
51&51667.7&0.12&$      8.19\pm      0.09$&$      1.48\pm      0.09$&&$      2.36\pm      0.02$&$     0.126\pm     0.003$&&$     0.327\pm     0.034$&$      2.15\pm      0.11$&0.7/66&SPL&\\
52&51668.8&0.12&$      8.22\pm      0.09$&$      1.46\pm      0.09$&&$      2.34\pm      0.02$&$     0.117\pm     0.003$&&$     0.322\pm     0.033$&$      2.08\pm      0.10$&1.0/66&SPL&\\
53&51669.2&0.12&$      8.06\pm      0.08$&$      1.61\pm      0.10$&&$      2.38\pm      0.02$&$     0.109\pm     0.003$&&$     0.228\pm     0.039$&$      2.38\pm      0.13$&1.1/66&SPL&\\
54&51670.6&0.10&$      8.09\pm      0.08$&$      1.58\pm      0.09$&&$      2.35\pm      0.02$&$     0.143\pm     0.003$&&$     0.323\pm     0.039$&$      1.85\pm      0.11$&0.8/66&SPL&\\
55&51670.8&0.11&$      8.29\pm      0.08$&$      1.56\pm      0.09$&&$      2.33\pm      0.02$&$     0.113\pm     0.003$&&$     0.314\pm     0.034$&$      1.95\pm      0.10$&0.8/66&INT&\\
56&51671.4&0.11&$      8.03\pm      0.09$&$      1.70\pm      0.10$&&$      2.35\pm      0.02$&$     0.108\pm     0.003$&&$     0.234\pm     0.044$&$      2.14\pm      0.13$&1.1/66&SPL&\\
57&51672.4&0.10&$      8.22\pm      0.09$&$      1.66\pm      0.09$&&$      2.31\pm      0.02$&$     0.121\pm     0.003$&&$     0.318\pm     0.036$&$      1.75\pm      0.10$&0.8/66&SPL&\\
58&51673.0&0.10&$      8.24\pm      0.08$&$      1.52\pm      0.07$&&$      2.39\pm      0.01$&$     0.205\pm     0.005$&&$     0.351\pm     0.055$&$      1.71\pm      0.14$&0.7/66&INT&\\
59&51673.4&0.10&$      8.14\pm      0.08$&$      1.58\pm      0.08$&&$      2.42\pm      0.01$&$     0.196\pm     0.004$&&$     0.283\pm     0.070$&$      1.84\pm      0.17$&0.8/66&INT&\\
60&51674.7&0.09&$      8.07\pm      0.10$&$      1.39\pm      0.08$&&$      2.31\pm      0.02$&$     0.224\pm     0.006$&&$     0.374\pm     0.076$&$      1.48\pm      0.16$&0.6/66&INT&\\

 \enddata

 \tablecomments{ \\
   1.  Reported error estimates are symmetric $1\sigma$ statistical uncertainties.\\
   2.  $M, i,$ and $D$ are frozen at their fiducial values.  }

\end{deluxetable}

\clearpage

\section{Continuum Fitting: Towards a Self-Consistent Disc + Reflection Model}\label{section:self-consistent}

In the previous section, we used the empirical model \smedge\ to crudely
account for a prominent spectral feature in the reflection component,
namely, the broad $K$-edge of iron.  We now consider a more
physically-motivated treatment of the full reflection spectrum, which is
generated by that portion of the power-law flux that strikes the
accretion disc \citep{rossfabian93}.  We first consider a generalised
version of \simpl that is more appropriate to the problem at hand.  We
then examine two reflection models, \ireflect, and \reflionx, concluding
that the former model is better for CF fitting, while the latter model
is better for fitting the profile of the Fe-\ka\ line (which is
considered in \S~\ref{reflection}).  As we describe below, there is
presently no unified reflection model that is well-suited to both
approaches of measuring spin.

\subsection{A Variant of the Power-Law Model \simpl}\label{subsec:simplr}

As in \S~\ref{section:analysis}, the core of our model consists of
\kerrbbtwo and \simplb.  However, we now introduce a modified version
of \simpl that is appropriate when including a separate and additive
reflection component.  This model, \simplr, is a generalisation of
\simpl that covers the two limiting cases described by equations 1 \&
2 in \citet{Steiner_simpl}, and applies to intermediate cases as well:

\begin{eqnarray}\label{eq:simplr}
\lefteqn{\nout(E)dE = (1-\fsc)\nin(E)dE } \nonumber \\
& &  {  } \qquad + (\fsc/x) \left[\int^{E_{\rm max}}_{E_{\rm min}}\nin(E_0)G(E;E_0)dE_0\right]dE.
\end{eqnarray}

Here, $\nin(E)$ and $\nout(E)$ are the seed input and model output
photon number densities at energy $(E)$.  Again, the normalisation
constant $\fsc$ is the fraction of photons directed into a power law
with photon index $\Gamma$, and $G(E;E_0)$ is the distribution function
of the output power law (see \citealt{Steiner_simpl, Ebisawa_1999}).
The one new parameter is $x$, which determines the fraction of the
power-law photons that strike the disc.  {\em These are the photons
which will be considered in modeling the reflection component.}

The standard version of \simplb, which was used in the preceding
section, assumes that none of the Compton-scattered photons strike the
disc, or adopting an equivalent interpretation, that reflection acts
like a perfect mirror with no absorption.  This corresponds to the
limiting case $x = 1$, which is described by Equation~1 in
\citet{Steiner_simpl}.  In the opposite limit, $x=2$, half of the
scattered photons are redirected downward, illuminating the disc,
while failing to reach an observer at infinity, As they encounter the
disc atmosphere, the returning photons are absorbed and reprocessed,
thereby generating the reflection component.  This limit corresponds
to Equation~2 in \citet{Steiner_simpl}.

The variant \simplr\ (Eq.~\ref{eq:simplr}) generalises this dichotomy,
making it possible to treat separately the reprocessed emission coming
from the illuminated disc via the tunable parameter $x$. This allows
one to model a corona quite generally.  The quantity $x-1$ describes
the solid angle subtended by the disc from the perspective of the
corona in units of 2$\pi$, which we refer to as a covering factor.  In
this paper, we assume that the geometry of the corona is a
disc-hugging slab with a covering factor of unity ($x = 2$); thus,
half the photons escape the system and half strike the disc.  As
shorthand, we will refer to the portion of the Compton component
produced by \simplr\ which irradiates the disc as \simplC\ (e.g., the
second term on the right-hand side of Eq.~\ref{eq:simplr} multiplied
by the covering factor).

\subsection{Results II: Continuum Fitting using \ireflect\ and \reflionx}\label{subsec:ireflect}

We first consider the model \ireflect\ \citep{PEXRAV}, which computes
the reflected spectrum (including scattering and edge absorption, but
excluding line fluorescence) generated in an ionised disc atmosphere
that is illuminated by an arbitrary external spectrum.  We convolve
the disc-illuminating component \simplC\ with \ireflect\ and isolate
the reflected component by setting the parameter {\it rel\_refl} to
-1.  (Our model implicitly assumes that the observed and illuminating
power-law spectra are identical.)  The ionisation parameter $\xi
\equiv L/nR^2$ is initially set to $10^4$ and allowed to vary freely
from $10^1-10^5$, while the characteristic disc temperature is fixed
to $T_{\rm disc} = 5 \times 10^6$ K and the metallicity is assumed to
be solar.  Fe-\ka\ emission is included separately in an approximate
fashion as an intrinsically narrow Gaussian line centred at a
restframe energy of 6.5~\keV.  This composite reflection component is
then convolved with the relativistic smearing kernel \kerrconv\
\citep{Brenneman_Reynolds} with the radial emissivity index $q$
fixed at the best-fitting \rxte value $q = 2.5$ (see \S~\ref{reflection}).
The complete model, which is comprised of an accretion-disc and a
power-law component, is: {\sc crabcor$\times$tbabs(\simplr$\otimes$\kerrbbtwo + kerrconv$\otimes$(ireflect$\otimes$\simplC\ + gauss))}.

The primary limitation of this model (referred to hereafter as Model~I)
is that although edges are included, the fluorescent line features
(e.g., \citealt{Garcia_Kallman_2010}), apart from Fe-\ka, are missing.
Also, the strength of the Fe-\ka\ feature should be tied to the depth of
the corresponding edge feature, but here that is not possible.  Below
and in \S~\ref{section:systematics}, we will demonstrate that these
shortcomings of Model~I have little effect on the CF spin results
because for our primary \gold\ spectra the reflected component is faint
compared to the dominant thermal component.  However, these issues are
of critical importance in estimating spin via the Fe-\ka\ line
(\S~\ref{reflection}).

We now consider a second reflection model, \reflionX~\citep{reflionx},
which we use as a replacement for (\ireflect$\otimes$\simplC+\gauss)
in Model~I given above.  We will refer to the new composite model as
Model~R.  In \reflionX, reflection is produced by a power-law spectrum
illuminating a cold slab of constant density.  The virtue of this
model is that it properly couples line emission to absorption, and it
also describes the full Fe-$K$ emission-line complex.  A major
drawback is that it is optimized for modeling AGN, which have cold
discs of lower density.  Consequently, in estimating spin via the
Fe-\ka\ method (\S~\ref{reflection}), we use a high-density variant of
\reflionX, \refbhb, which includes an intrinsic blackbody component
\citep{refbhb}.  Because the blackbody component is hardwired into
\refbhb, it can not presently be used with CF models.  We therefore
use \reflionX\ in concert with \kerrbbtwo for our CF analysis.

In addition to the temperature/density limitations of \reflionX\ just
mentioned, this model has additional shortcomings.  Of primary
importance, it requires that the illuminating spectrum have a simple
power law form.  This power law is not truncated at low energies and
its flux can rival or exceed the thermal flux, thereby leading to
unphysical results (see e.g., \citealt{Steiner_simpl}).  In addition, the
strength of the reflected component is not linked to the normalisation
of the illuminating spectrum, and so there is no way to ensure that
the Compton and reflection components are appropriately matched.
Nevertheless, we employ Model~R using \reflionX\ as a second-tier CF
model that gives us an independent check on the results obtained using
Model~I.

In summary, using a variant of \simpl and considering two reflection
models, we have progressed toward a model featuring a self-consistent
treatment of thermal disc emission, Compton scattering, and disc
reflection.  For estimating spin via the CF method, we favor \ireflect,
while for the Fe-\ka\ method we elect to use \reflionx\ and \refbhb~
(\S~\ref{reflection}).

%\subsection{Results II}\label{subsec:resultsII}

We now apply Model~I (\S~\ref{subsec:ireflect}) to our set of \rxte
spectra, while following the procedures described in
\S~\ref{subsec:resultsI}.  In this case, we find that only a total of
45 spectra (24 \gold\ and 21 \silver) meet our selection criteria
(\S~\ref{subsec:selection}), compared to the 60 selected using
Model~S.  Our fitting results for these 45 spectra are given in
Table~\ref{tab:ireflect}.

\begin{figure}
\centering{\includegraphics[clip=true, angle=90,width=8.75cm]{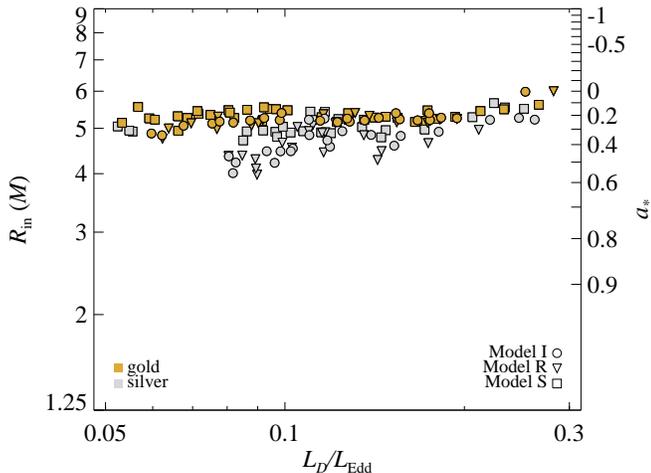}}
\caption{As in Figure~\ref{fig:lumspin}, we again plot $\Rin$ and
  $\spin$ versus luminosity, but we now show results for all three of
  the models discussed in the text.  The data for Model~S, which are
  repeated from Figure~\ref{fig:lumspin}, show the highest degree of
  internal consistency.}\label{fig:compare3}
\end{figure}

\begin{figure}
\centering{\includegraphics[clip=true, angle=90,width=8.95cm]{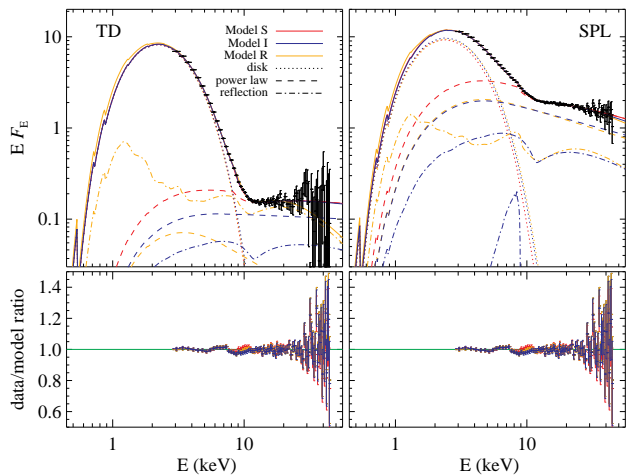}}
\caption{Model fits for a \gold\ spectrum ({\it left}) and a \silver\
  spectrum ({\it right}), which correspond respectively to observations
  made on MJD 51121.0 and MJD 51115.3 (see Tabs. \ref{tab:smedge} \&
  \ref{tab:ireflect}).~The models are differentiated by line colour and
  the individual components by line texture.~Note how much weaker the
  power-law component is for the \gold\ spectrum, and how closely all
  three models track the data. }\label{fig:modfit}
\end{figure}

For the \gold\ spectra, we find excellent agreement between the
results obtained using Model~I, $\spin = 0.23 \pm 0.06$, and Model~S,
$\spin = 0.20 \pm 0.04$.  This agreement is illustrated in
Figure~\ref{fig:compare3}, where we also show results for Model~R.  As
is apparent, considering only the \gold\ spectra, all three models are
in excellent agreement -- the mean values of $\rin$ are consistent
with one another to within $\approx2$ per~cent.  However, for the \silver\
spectra the mean values of $\rin$ are depressed for all three models,
by $\approx10$ per~cent for the self-consistent reflection models (which track
each other closely) and by only $\approx5$ per~cent for Model~S.
Interestingly, the primitive Model~S performs better than the
self-consistent reflection models by harmonizing the results obtained
from the two data sets and delivering the highest degree of internal
consistency.  Figure~\ref{fig:modfit} shows an overlay comparison of
the best-fitting results using the three models for two representative
spectra, one \gold\ and the other \silver.  The total unfolded spectra
and their components are plotted, as well as the data/model ratio. 
The key result of this section is that using our fiducial values of
$M$, $i$ and $D$ (\S~\ref{section:Intro}), all three models applied to
the {\em gold} spectra give the same low estimate
of~spin:~$0.15{\thinspace<}\spin{\thinspace<}0.35$.

%   GOLD:    MJD 51121.0  
%   SILVER:  MJD 51115.3

\section{Continuum Fitting: Error Analysis and Final Spin Result}\label{section:systematics}

In this section, we broadly consider three sources of observational
error, both systematic and statistical, which bear on our final estimate
of the spin.  In order of increasing importance, these are (1)
sensitivity to the details of the spectral models employed; (2) X-ray
flux calibration uncertainties; (3) and the uncertainties in the input
parameters $M$, $i$ and $D$.  We then perform a comprehensive analysis
that incorporates these uncertainties and arrive at our final CF
estimate of the spin of the black hole.  In the following, we present an
overview; for details, see Appendix~\ref{append:systematics}.

\begin{itemize}

\item {\it Sensitivity to X-ray spectral models.}  In order to make this
  assessment, we determine the change in $\rin$ when varying a single
  model component or a single one of the parameter settings.
  Table~\ref{tab:systematics} gives the mean fractional change in $\rin$
  for the \gold\ data sample that arises from changing either a model
  parameter (rows P1-9) or a model component (M1-5).  The `Change'
  column lists either the new value adopted for the parameter or it
  describes the change made to the model component (where DISC, PL, and
  REFL refer to the accretion disc, power law, and reflection
  components).  The third and fourth columns respectively list the
  fractional changes in $\rin$ for Model~I and Model~S.  The largest
  tabulated uncertainty arises from the choice of the viscosity
  parameter: Using $\alpha = 0.01$ instead of the default value ($\alpha
  = 0.1$) decreases $\rin$ by $\approx 6$ per~cent for Model~I and 3 per~cent for
  Model~S.  Each of the other 13 changes considered affect $\rin$ by
  $<3$ per~cent for either model.
\end{itemize}

  \begin{deluxetable}{llcccccccccccccccccccc} 
  %\rotate 
  \tabletypesize{\scriptsize} 
  \tablecolumns{          22}
  \tablewidth{0pc}  
  \tablecaption{Model~I Continuum-Fitting Results}
  \tablehead{\colhead{N} & \colhead{MJD} & \colhead{$\frac{L_D}{L_{\tiny{\textrm{Edd}}}}$} &  \multicolumn{2}{c}{\simplr}  &  &  \multicolumn{2}{c}{\kerrbbtwo} & & \colhead{\ireflect} & & \colhead{\gauss}  & \colhead{$\chi ^2 _\nu $/DOF}  & \colhead{State} \\ \cline{4-5} \cline{7-8} \cline{10-10} \cline{12-12} \\  & &  &  \colhead{$\Gamma$}  &  \colhead{$f_{sc}$}  &  &  \colhead{$\spin$}  &  \colhead{$\Mdot (10^{18} {\rm g/s})$}  &  &  \colhead{$\xi {\rm (erg~cm/s)}$}  &  &  \colhead{$N {\rm(10^{-3}/cm^2/s)}$} } 
 \startdata  \label{tab:ireflect}
 
1&51117.4&0.18&$      2.04\pm      0.04$&$     0.018\pm     0.002$&&$     0.235\pm     0.021$&$      3.37\pm      0.11$&&$1522\pm1560$&&$       2.8\pm       2.7$&0.7/74&TD&\\
2&51119.0&0.16&$      2.07\pm      0.04$&$     0.017\pm     0.002$&&$     0.230\pm     0.021$&$      3.01\pm      0.09$&&$1215\pm1345$&&$       2.7\pm       2.0$&0.7/74&TD&\\
3&51121.0&0.14&$      2.06\pm      0.06$&$     0.006\pm     0.001$&&$     0.238\pm     0.017$&$      2.61\pm      0.07$&&$1967\pm2881$&&$       1.4\pm       1.1$&0.7/74&TD&\\
4&51124.7&0.11&$      2.07\pm      0.06$&$     0.012\pm     0.002$&&$     0.240\pm     0.022$&$      2.20\pm      0.07$&&$1179\pm2224$&&$       2.5\pm       1.5$&0.6/74&TD&\\
5&51128.6&0.10&$      2.09\pm      0.05$&$     0.020\pm     0.003$&&$     0.236\pm     0.024$&$      1.89\pm      0.07$&&$1101\pm1834$&&$       2.9\pm       1.5$&0.9/74&TD&\\
6&51130.5&0.09&$      2.15\pm      0.03$&$     0.023\pm     0.001$&&$     0.221\pm     0.018$&$      1.79\pm      0.05$&&$311.1\pm114.6$&&$       2.9\pm       0.9$&1.1/74&TD&\\
7&51132.5&0.08&$      2.07\pm      0.03$&$     0.018\pm     0.001$&&$     0.254\pm     0.030$&$      1.55\pm      0.07$&&$1976\pm1868$&&$       2.9\pm       1.2$&0.8/74&TD&\\
8&51134.5&0.08&$      2.15\pm      0.04$&$     0.013\pm     0.001$&&$     0.261\pm     0.028$&$      1.43\pm      0.06$&&$1935\pm2203$&&$       1.9\pm       0.8$&0.8/74&TD&\\
9&51136.9&0.07&$      2.08\pm      0.21$&$     0.013\pm     0.008$&&$     0.276\pm     0.031$&$      1.26\pm      0.07$&&$732.7\pm5296$&&$       2.0\pm       1.0$&0.9/74&TD&\\
10&51157.6&0.12&$      2.63\pm      0.70$&$     0.004\pm     0.007$&&$     0.251\pm     0.020$&$      2.33\pm      0.08$&&$606.7\pm10500$&&$       0.5\pm       0.8$&0.8/74&TD&\\
11&51160.3&0.17&$      1.97\pm      0.32$&$     0.001\pm     0.001$&&$     0.237\pm     0.017$&$      3.21\pm      0.08$&&$50600\pm741400$&&$       0.0\pm       1.5$&0.7/74&TD&\\
12&51163.2&0.25&$      2.66\pm      1.29$&$     0.001\pm     0.003$&&$     0.005\pm     0.028$&$      5.62\pm      0.17$&&$10\pm3988$&&$       3.4\pm       3.0$&1.7/74&TD&\\
13&51260.6&0.19&$      1.97\pm      0.04$&$     0.015\pm     0.001$&&$     0.222\pm     0.021$&$      3.77\pm      0.11$&&$5823\pm4795$&&$       4.5\pm       3.0$&0.4/66&TD&\\
14&51261.8&0.18&$      2.04\pm      0.04$&$     0.018\pm     0.002$&&$     0.218\pm     0.022$&$      3.56\pm      0.12$&&$1527\pm1575$&&$       2.4\pm       2.9$&0.7/66&TD&\\
15&51263.1&0.17&$      2.07\pm      0.04$&$     0.020\pm     0.002$&&$     0.184\pm     0.034$&$      3.48\pm      0.15$&&$1869\pm1752$&&$       4.7\pm       3.0$&0.5/66&TD&\\
16&51264.8&0.15&$      1.97\pm      0.03$&$     0.014\pm     0.001$&&$     0.184\pm     0.032$&$      3.06\pm      0.12$&&$5574\pm3975$&&$       7.2\pm       1.9$&0.8/66&TD&\\
17&51265.6&0.14&$      2.10\pm      0.08$&$     0.022\pm     0.004$&&$     0.219\pm     0.020$&$      2.79\pm      0.09$&&$547\pm1228$&&$       3.3\pm       1.5$&0.7/66&TD&\\
18&51266.9&0.13&$      2.08\pm      0.23$&$     0.014\pm     0.008$&&$     0.247\pm     0.027$&$      2.45\pm      0.12$&&$672.7\pm4994$&&$       2.6\pm       2.0$&0.5/66&TD&\\
19&51267.6&0.13&$      2.08\pm      0.17$&$     0.022\pm     0.009$&&$     0.225\pm     0.027$&$      2.47\pm      0.13$&&$697.8\pm3756$&&$       4.9\pm       2.6$&0.8/66&TD&\\
20&51273.6&0.10&$      1.95\pm      0.03$&$     0.015\pm     0.001$&&$     0.183\pm     0.040$&$      1.97\pm      0.10$&&$10000\pm8847$&&$       8.1\pm       1.4$&1.9/66&TD&\\
21&51274.5&0.09&$      2.04\pm      0.10$&$     0.018\pm     0.005$&&$     0.238\pm     0.028$&$      1.68\pm      0.08$&&$827.3\pm2862$&&$       3.1\pm       1.6$&0.8/66&TD&\\
22&51276.3&0.08&$      1.99\pm      0.19$&$     0.010\pm     0.004$&&$     0.246\pm     0.028$&$      1.48\pm      0.07$&&$516.7\pm2565$&&$       2.6\pm       0.7$&0.7/66&TD&\\
23&51278.7&0.06&$      1.92\pm      0.06$&$     0.011\pm     0.001$&&$     0.341\pm     0.031$&$      1.11\pm      0.06$&&$23140\pm53800$&&$       3.6\pm       1.2$&0.8/66&TD&\\
24&51279.6&0.06&$      1.99\pm      0.31$&$     0.013\pm     0.011$&&$     0.329\pm     0.022$&$      1.07\pm      0.05$&&$677.9\pm7036$&&$       2.1\pm       1.1$&0.8/66&TD&\\
\hline
25&51110.3&0.26&$      2.45\pm      0.02$&$     0.109\pm     0.005$&&$     0.233\pm     0.033$&$      5.06\pm      0.20$&&$4879\pm2066$&&$      16.8\pm       9.5$&0.8/74&INT&\\
26&51111.6&0.25&$      2.39\pm      0.03$&$     0.113\pm     0.006$&&$     0.220\pm     0.040$&$      4.80\pm      0.23$&&$7147\pm4072$&&$      24.1\pm      11.0$&0.7/74&INT&\\
27&51113.7&0.22&$      2.43\pm      0.03$&$     0.108\pm     0.007$&&$     0.234\pm     0.045$&$      4.25\pm      0.24$&&$2331\pm1531$&&$      16.7\pm       9.4$&0.9/74&INT&\\
28&51115.3&0.18&$      2.39\pm      0.07$&$     0.086\pm     0.011$&&$     0.317\pm     0.041$&$      3.28\pm      0.26$&&$875.9\pm1712$&&$       9.0\pm       6.9$&0.7/74&SPL&\\
29&51126.6&0.11&$      2.14\pm      0.03$&$     0.025\pm     0.001$&&$     0.233\pm     0.024$&$      2.11\pm      0.08$&&$10000\pm6716$&&$       8.8\pm       1.7$&1.5/74&INT&\\
30&51269.7&0.12&$      2.10\pm      0.03$&$     0.022\pm     0.001$&&$     0.225\pm     0.025$&$      2.26\pm      0.08$&&$10000\pm8200$&&$       9.1\pm       1.9$&1.4/66&SPL&\\
31&51270.8&0.08&$      2.17\pm      0.03$&$     0.053\pm     0.003$&&$     0.502\pm     0.025$&$      1.28\pm      0.06$&&$10000\pm6763$&&$      10.2\pm       2.5$&1.6/66&SPL&\\
32&51271.4&0.08&$      2.20\pm      0.03$&$     0.059\pm     0.003$&&$     0.469\pm     0.044$&$      1.29\pm      0.10$&&$10000\pm7370$&&$       9.5\pm       3.0$&1.2/66&SPL&\\
33&51664.4&0.16&$      2.40\pm      0.04$&$     0.090\pm     0.005$&&$     0.345\pm     0.029$&$      2.78\pm      0.12$&&$511.2\pm558.9$&&$      10.2\pm       3.5$&0.6/66&SPL&\\
34&51664.7&0.15&$      2.40\pm      0.05$&$     0.097\pm     0.007$&&$     0.407\pm     0.024$&$      2.58\pm      0.13$&&$568.4\pm836.3$&&$      10.4\pm       3.8$&0.9/66&SPL&\\
35&51665.4&0.14&$      2.37\pm      0.13$&$     0.079\pm     0.019$&&$     0.339\pm     0.055$&$      2.49\pm      0.31$&&$745.6\pm3070$&&$       7.6\pm       6.3$&0.6/66&SPL&\\
36&51667.7&0.12&$      2.29\pm      0.03$&$     0.067\pm     0.005$&&$     0.412\pm     0.029$&$      2.00\pm      0.11$&&$1533\pm1302$&&$       8.3\pm       3.9$&0.7/66&SPL&\\
37&51668.8&0.12&$      2.32\pm      0.04$&$     0.068\pm     0.005$&&$     0.374\pm     0.034$&$      2.04\pm      0.11$&&$515.3\pm602.9$&&$       6.2\pm       2.1$&0.8/66&SPL&\\
38&51669.2&0.12&$      2.35\pm      0.02$&$     0.063\pm     0.002$&&$     0.314\pm     0.023$&$      2.27\pm      0.08$&&$232.8\pm91.73$&&$       7.6\pm       2.5$&0.9/66&SPL&\\
39&51670.6&0.10&$      2.30\pm      0.03$&$     0.079\pm     0.004$&&$     0.423\pm     0.023$&$      1.72\pm      0.07$&&$455.3\pm353.1$&&$       8.0\pm       2.3$&0.7/66&SPL&\\
40&51670.8&0.10&$      2.25\pm      0.03$&$     0.060\pm     0.004$&&$     0.439\pm     0.030$&$      1.68\pm      0.09$&&$1661\pm1319$&&$       7.2\pm       3.2$&0.7/66&INT&\\
41&51671.4&0.11&$      2.29\pm      0.13$&$     0.058\pm     0.016$&&$     0.340\pm     0.036$&$      1.96\pm      0.17$&&$646.3\pm2684$&&$       7.3\pm       2.5$&0.9/66&SPL&\\
42&51672.4&0.09&$      2.25\pm      0.10$&$     0.067\pm     0.014$&&$     0.439\pm     0.040$&$      1.53\pm      0.15$&&$818.8\pm2652$&&$       7.4\pm       4.4$&0.7/66&SPL&\\
43&51673.0&0.10&$      2.33\pm      0.07$&$     0.109\pm     0.013$&&$     0.504\pm     0.045$&$      1.49\pm      0.16$&&$947.3\pm1858$&&$      10.1\pm       5.7$&0.6/66&INT&\\
44&51673.4&0.10&$      2.36\pm      0.09$&$     0.103\pm     0.017$&&$     0.440\pm     0.060$&$      1.61\pm      0.23$&&$860.7\pm2522$&&$      10.1\pm       6.1$&0.8/66&INT&\\
45&51674.7&0.08&$      2.22\pm      0.03$&$     0.113\pm     0.006$&&$     0.556\pm     0.046$&$      1.20\pm      0.11$&&$2605\pm1802$&&$      11.3\pm       5.6$&0.6/66&INT&\\
     
 \enddata 

 \tablecomments{ \\
   $^1$ Reported error estimates are symmetric $1\sigma$ statistical uncertainties.\\
   $^2$ $M, i,$ and $D$ are frozen at their fiducial values.  }

\end{deluxetable}

\clearpage

\begin{itemize}

\item {\it Flux calibration.}  The problem of flux calibration is
endemic to X-ray astronomy.  The Crab spectrum, as determined by
\citet{Toor_Seward}, is the widely-adopted standard that we have
consistently used in our work.  Uncertainties in the normalisation of
this spectrum have recently been considered by \citet{Weisskopf:crab}.
Using their Figure~1 as a guide, we adopt a generous $\pm 10$ per~cent
uncertainty in our overall flux calibration, which corresponds to a
$5$ per~cent uncertainty in $\rin$.

\item {\it Uncertainties in $M$, $i$ and $D$.}  As in our earlier work
  (e.g., \citealt{Liu_m33x7, Gou_2009, Gou_2010}), we sample the allowed
  parameter space assuming Gaussian errors (except here for $D$, we use
  an asymmetric Gaussian).  The sampling is performed using 42,500
  triplets of $M$, $i$, and $D$, which are distributed in a uniform grid
  throughout the parameter space.  At each point in the grid, the
  complete \rxte data set is analysed with Model~S, and the selection
  criteria given in \S~\ref{section:analysis} are separately applied to
  the results.  Folding all of the runs together, a composite
  distribution based on all of the selected spectra is obtained, where
  we have additionally weighted over the set of possible dynamical
  models (see Table~1 in \citealt{Orosz_Steiner_2010}).

  In conducting this analysis, we have included the robust no-eclipse
  constraint, $i < 82\degr$.  We have further required that during the
  TD-state plateau phase (days 105--182; Figure~\ref{fig:state-lc}) the
  disc luminosity not exceed 85 per~cent of $\ledd$ (the actual Eddington limit
  for disc geometry; see Section 6.1 in \citealt{McClintock_2006}).
  Lastly, we also require that the disc luminosity during the thermal
  plateau phase be greater than 10 per~cent of $\ledd$, or else the full sample
  of TD data would extend downward in luminosity to the unreasonably low
  value of $\lesssim 0.1$ per~cent $\ledd$.
  
\end{itemize}

In the analysis described above, we have used the default value of the
viscosity parameter, $\alpha = 0.1$.  Because $\alpha$ is the major
source of uncertainty considered in Table~\ref{tab:systematics}, we
have repeated the analysis just described using $\alpha = 0.01$ and
combined the two distributions, weighting them equally.  We combine
all other errors in Table~\ref{tab:systematics}, yielding an ensemble
value of $\approx 4.2$ per~cent.  Finally, we add in quadrature the 5
per~cent error in the absolute flux calibration and arrive at our net
error of 6.5 per~cent.  The effect of this uncertainty on our
measurement of spin is incorporated by running a boxcar smoothing
kernel (with a 13 per~cent full width) over the distribution for
$\rin$.

\begin{figure}
{\includegraphics[clip=true,angle=90,width=8.7cm]{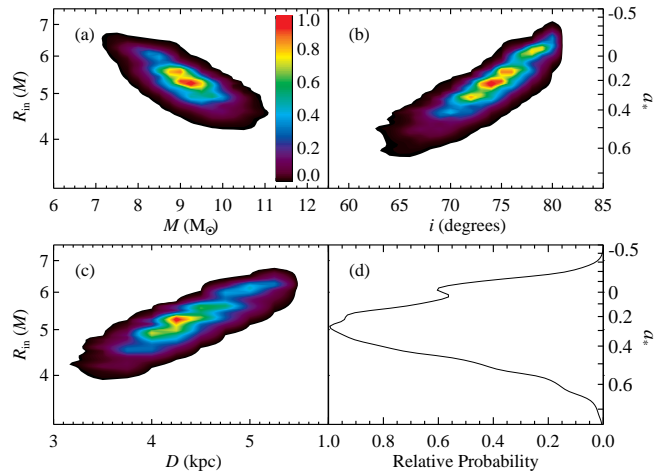}}
\caption{Probability contours for the relationship between $\rin$/spin
  and $M$, $i$, and $D$.  Each of the first three panels ({\it a-c})
  shows variation for a single parameter; the other two parameters
  have been fixed at their best values.  The orientations of the
  probability ellipsoids show that spin is positively correlated with
  $M$ and negatively correlated with both $i$ and $D$.  In panel ({\it
    d}), a combined probability distribution for the case $\alpha =
  0.1$ is shown (arbitrarily scaled) with variation in $M$, $i$, and
  $D$ folded together.}\label{fig:CF-correlations}
\end{figure}

The dominant source of error is the observational uncertainties in
$M$, $i$ and $D$, which in turn are largely due to the uncertainties
associated with modeling the optical/NIR light curves
\citep{Orosz_Steiner_2010}.  Figure~\ref{fig:CF-correlations} shows
the dependence of $\rin$/spin on these model parameters.  Here, using
the results of the grid analysis described above, we vary one of the
three parameters, fixing the other two at the their best values.  The
strong correlations between spin and inclination, and between spin and
distance, demonstrate the degree to which measurement errors in these
quantities contribute to the uncertainty in spin.  Together, errors in
$M$, $i$, and $D$ account for $\Delta\spin \approx 0.25$
($\Delta\rin/\rin \approx 0.2$) for the 90 per~cent confidence interval.  The
contribution due to the inclination is sizable, $\sim 11$ per~cent, because
its uncertainty and its value are large ($\lesssim 4$ degrees and 74.7
degrees, respectively).  The uncertainty in $D$ likewise contributes
$\sim 11$ per~cent, while uncertainties in the mass affect the spin only at
the $\sim 7$ per~cent level.

\begin{table}
\begin{centering}
\caption{ \label{tab:systematics} 
Systematic Changes to the Model} 
\begin{tabular}{l l r r r }
\hline
TD & Change & \multicolumn{2}{c}{$\bar{\Delta\rin}(\%)$} \\
\cline{3-4} \\
 & & Model~I & Model~S \\
\hline \hline
   P1 & $\alpha = 0.01$                  &    -5.98  &  -2.90  \\
   P2 & $\nh = 6 \times 10^{21}\pcmsq$    &    -2.90  &  -1.94  \\
   P3 & $\nh = 10 \times 10^{21}\pcmsq$   &     0.84  &   2.84   \\
   P4 & $ x  = 1.5$                       &    -1.13  & \nodata  \\
   P5 & $ x  = 1.1$                       &    -1.26  & \nodata  \\
   P6 & $ q$ and $x $ free                &     0.21  & \nodata  \\
   P7 & $T_{\rm disk} = 10^6$\ K           &    -0.09  & \nodata  \\
   P8 & $ W_{\rm Edge} = 3.5~\keV $        &   \nodata &  -0.42    \\
   P9 & $ W_{\rm Edge} = 14~\keV  $        &   \nodata &   0.41    \\
   \hline
   M1 & DISK: \bhspec                    &     1.30  &   2.88    \\
   M2 & PL: falloff with $kT_e$           &    -1.53  &  -0.66    \\
   M3 & PL: down-scattering set           &    -0.27 &   0.07    \\
   \hline
   M4 &  REFL: \smedge                   &     1.70  &   \nodata \\
   M5 &  REFL: \reflionX                 &    -1.00  &   -2.65   \\
   \hline
\end{tabular}
\end{centering}
\end{table}
 %\clearpage

\begin{figure}
{\includegraphics[angle=90,width=9.9cm]{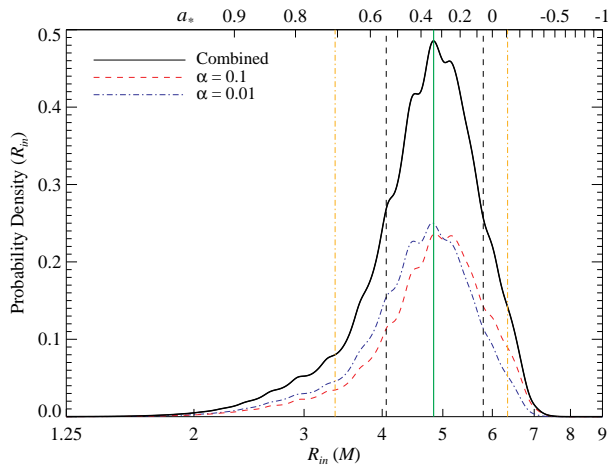}}
\caption{Composite probability density for $\Rin$ and $\spin$ which
  takes into account both systematic and statistical errors including:
  uncertainties in distance, black hole mass and inclination; the
  spectral model; and the uncertainty in the absolute flux
  calibration.  The net probability distribution is a combination of
  the individual distributions for two values of $\alpha$.  The
  contribution from each integrates to 50 per~cent probability and is shown
  for $\alpha = 0.01 $ (blue dash-dotted line) and $\alpha = 0.1$ (red
  dashed line).  The 90 per~cent confidence limits for the combined
  distribution are shown as yellow vertical lines, the 1$\sigma$
  limits as vertical black lines, and the most probable spin is marked
  with a green line.  We conclude that the spin is moderate: $-0.11 <
  \spin < 0.71$ (90 per~cent confidence).}\label{fig:spin_mid_gauss}
\end{figure}

After folding together all sources of error, the resulting probability
distribution is shown in Figure~\ref{fig:spin_mid_gauss}, with $\Rin$
and $\spin$ displayed respectively on the bottom and top axes.  The
green vertical line identifies the most probable spin, $\spin = 0.34$,
and the yellow lines indicate the 90 per~cent confidence interval, which
extends from -0.11 to 0.71.  From an inspection of this distribution
function, we conclude that the black hole is unlikely to be in a
retrograde configuration (only $\sim 11.2$ per~cent probability).  Of greater
importance, we conclude that {\it the spin is not high}.  For example,
the probability that the CF spin exceeds 0.9 is less than 0.4 per~cent, a
surprising result for a black hole that has produced superluminal
jets.

%%%%%%%%%%%%%%%%%%%%%%%%%%%%%%%%%%%%%%%%%%%%%%%%%%%%%%%%%%
%%%%%%%%%%%%%                     %%%%%%%%%%%%%%%%%%%%%%%%
%%%%%%%%%        START    RUBENS       %%%%%%%%%%%%%%%%%%%
%%%%%%%%%%%%%                     %%%%%%%%%%%%%%%%%%%%%%%%
%%%%%%%%%%%%%%%%%%%%%%%%%%%%%%%%%%%%%%%%%%%%%%%%%%%%%%%%%%

\section{Spin from Reflection Features}
\label{reflection}

In the previous section, we concluded that the spin parameter has a
low or intermediate value.  This result is based on our CF analysis of
many \rxte\ spectra, which were obtained primarily in the TD state.
In what follows, we first analyse intermediate-state spectra of \j\
obtained with the \asca X-ray observatory; we then supplement this
analysis using a sample of \rxte\ spectra, also obtained in the
intermediate state.  We fit the \asca GIS-2 and GIS-3 data
simultaneously, using a floating normalisation constant to allow for
cross-calibration uncertainties.  Our work differs from earlier
analyses of these same data by others (e.g., see
\citealt{Miller_2005_j1550, gierlinskidone03, Miller_2009}): Our focus
is on a detailed analysis of the reflection component, rather than on
a precise model of the overall continuum.  We begin by setting all the
physical parameters of the binary to the best-estimate values
presented in \citet{Orosz_Steiner_2010}.

In addition to the soft disc and hard power-law components seen in the
TD- and intermediate-state spectra of black hole binaries, a broad
emission line at $\sim 6.4$~\kev\ is also often present (see
e.g. \citealt{miller07review}).  This line feature is merely the most
prominent reflection signature that arises as hard emission from the
corona irradiates the cooler disc (\citealt{reflionx}).  In the
vicinity of a black hole, the iron-\ka\ line shape and other
reflection features are distorted by various relativistic effects
(\citealt{Fabian_1989, laor}). The spin parameter can be constrained
by modeling these features because their shape depends on how far the
disc extends down into the gravitational potential well (see
\S~\ref{section:Intro}), the key assumption again being that this
extent is set by the radius of the ISCO.

\subsection{Phenomenological Models -- \asca}
\label{simple}

\begin{figure}
{\includegraphics[angle=0,width=7.5cm, clip=true]{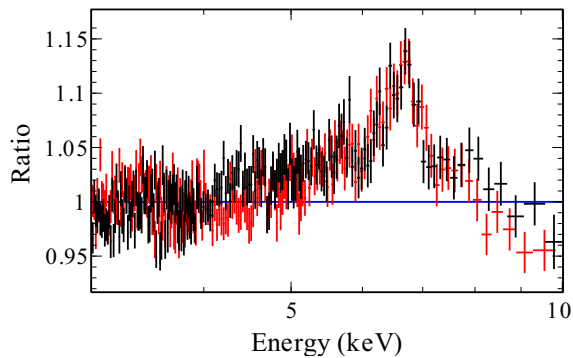} }
\caption{Data/model ratio for a phenomenological continuum model
  consisting of a thermal disc and a Compton component. \asca GIS-2
  and GIS-3 spectra are shown in black and red respectively. The data
  were fitted jointly in the 1--4 and 7--10~\kev\ energy range. The
  residuals in the 4--7~\kev\ band show the relativistic nature
  (asymmetry and broadness) of the iron-emission line profile. The
  data have been rebinned for plotting purposes. The data/model ratio
  for the full energy range is shown in Fig.~\ref{fig_ratiopheno}.}
\label{fig_ratio2pl}
\end{figure}

In order to highlight the relativistic nature of the line profile in
the \asca spectra, we start by modeling the 1--4~\kev\ and 7--10~\kev\
continua with a combination of a disc blackbody (described by the
\xspec model \diskbb {\footnote {This model characterises the thermal
    emission using only two parameters -- the flux normalisation and a
    colour temperature.  Here, we use this very approximate model of
    the continuum (compare \S~\ref{section:analysis}) because of its
    simplicity in \emph{phenomenologically} describing the thermal
    continuum.}}  of \citealt{diskbb}) and the Comptonisation model
\simplb.  The neutral hydrogen column was initially fixed at $\nh =
8\times10^{21}\pcmsq$ as per \citet{Miller_2003_j1550}, which resulted
in a poor fit to the continuum with $\chi^2/\nu = 2172.3/1002$.
Allowing the column density to vary resulted in a significant
improvement to the fit with $\chi^2/\nu = 1367.5/1001$ for $\nh =
5.4\pm0.1\times10^{21}\pcmsq$.  The total neutral hydrogen column
density in the line of sight to \j, which was determined using the
\chandra transmission grating, is not expected to vary
(\citealt*{miller_nhpaper}).  However, the differing values of $\nh$
can be reasonably attributed to differences in the calibrations of the
\chandra and \asca detectors.  Furthermore, allowing $\nh$ to differ
between the two GIS spectra further improves the fit: $\Delta\chi^2 =
-47.5$ for one less degree of freedom with a difference in $\nh$ of
$<5$ per~cent.  Figure~\ref{fig_ratio2pl} shows the line spectrum obtained by
modeling the continuum as described above.  The asymmetric and broad
residual feature in the 4--7~\kev\ band has the appearance one expects
for fluorescent disc-line emission arising near a black hole.

\begin{figure}
\centering{\includegraphics[angle=0, width=8.75cm]{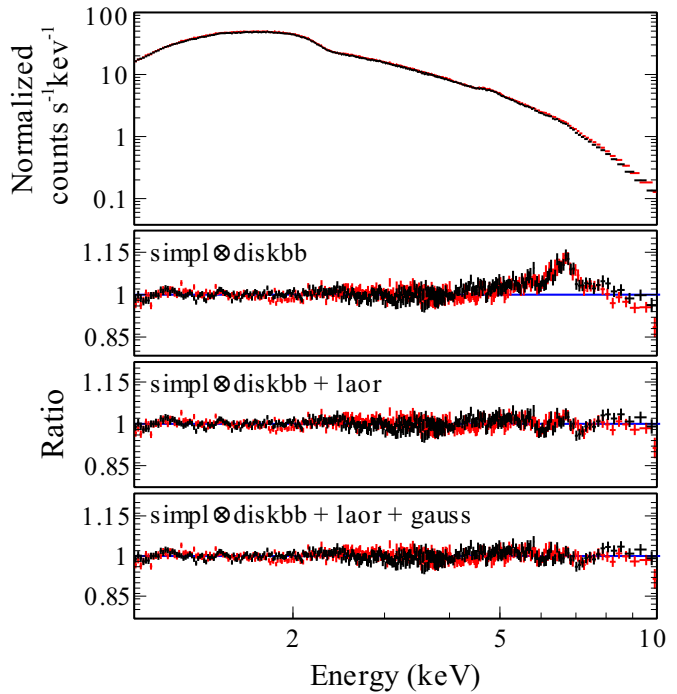}}
\caption{{\it (top:)} The \asca spectra.  Below are plots of the ratio
  of the data to a phenomenological continuum model consisting of
  thermal-disc and Compton components plus a \laor\ line; in the
  bottommost panel, a narrow Gaussian line has been added to the
  model. }
  \label{fig_ratiopheno}
\end{figure}

We provide a physical description of the Fe line by first modeling the
residuals seen in Figure~\ref{fig_ratio2pl} using the \laor\ model
\citep{laor} and fitting for the inner radius \rrin\ and the
power-law index $q$ of the emissivity profile, which is described by a
power-law of the form $\epsilon(r) \propto r^{\it -q}$.  The outer disc
radius is fixed at the maximum allowed value of 400~\rrg\ (\rrg$\equiv
GM/c^2$), and the disc inclination is constrained to be approximately
1$\sigma$ from the adopted value of \citet{Orosz_Steiner_2010}
(i.e. between 71 and 78 degrees).  The line energy is constrained
between 6.4--6.97~\kev.  The fit achieved by including the \laor\
component, shown in Figure~\ref{fig_ratiopheno}, results in $\chi^2/\nu
= 1848.0/1501$, an improvement of $\Delta\chi^2=-416$ for 5 fewer
degrees of freedom (compared to the best-fitting continuum model with no
line feature).  The best fit parameters for this model are detailed in
Table~\ref{table} (Model~1). 

It can be seen from the ratio plot shown in
Figure~\ref{fig_ratiopheno} that this simple, heuristic model,
although mostly adequate, does not provide a detailed description of
all the features present in the 6--8~\kev\ range. Adding a narrow
Gaussian line at $\approx 6.7$~\kev\ only marginally improves the
fit{\footnote{This feature was previously associated
    (\citealt{tomsick01j1550}) with emission from the Galactic ridge
    (\citealt{valinia98}).}}  ($\Delta\chi^2=-14.4$ for 2 fewer
degrees of freedom), with evidence for additional residuals, which are
possibly associated with Fe-$K$-shell absorption edges in partially
ionised material (\citealt{rossfabian93, rossfabianbrandt96}). Such
features are usually present at $\approx 7.1$~\kev\ in TD-state
spectra of black hole binaries (\citealt{done92, reisgx}). In order to
properly account for the panorama of features associated with the
reprocessing of radiation in the accretion disc, we now consider
complete reflection models.

\subsection{Reflection Analysis -- \asca}
\label{full_refl}

\begin{figure}
\centering{\includegraphics[angle=0, width=8.5cm]{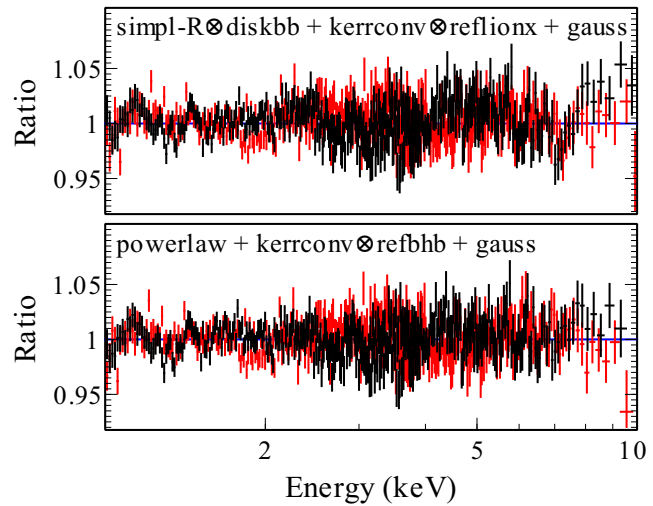}}
\caption{Data/model ratio for ({\it top:}) the \reflionx\ model
  together with a separate thermal emission and Compton
  component. ({\it bottom:}) Self-consistent thermal emission and
  reflection (\refbhb) together with a power-law component.  }
\label{fig_ratio_ref}
\end{figure}

We replace the \laor\ component with \reflionx\ (\citealt{reflionx}),
which describes the spectrum reflected from an optically-thick and
cold atmosphere of constant density that is illuminated by a power-law
spectrum (\S~\ref{section:self-consistent}).  The parameters of the
model are the iron abundance (set to Solar), photon index of the
illuminating power law, ionisation parameter, and normalisation.  The
gravitational and Doppler effects are accounted for using the fully
relativistic convolution model \kerrconv\
(\citealt{Brenneman_Reynolds}), which includes black hole spin as a
fit parameter.  The power-law indexes of \reflionx\ and the Compton
component (\simplr) are tied, and, as before, we constrain the
inclination to lie between 71 and 78 degrees and include a narrow
Gaussian line at $\approx6.7$~\kev.  The model results in a good and
improved fit to the data with $\chi^{2}/\nu = 1752.3/1499$ (Model~2 in
Table~\ref{table}); however it still does not fully account for the
reflection features, with residuals present at $\approx 7$~\kev\ (top
panel of Fig.~\ref{fig_ratio_ref}).

Although the scattered fraction for this spectrum is high, $\fsc >
50$ per~cent, and the CF method is not applicable, we nevertheless
investigated the effect of switching the continuum model from \diskbb\
to \kerrbb (Model~3), with the mass, distance and inclination frozen
at their nominal values.  This change produced insignificant
differences in the fit parameters (Table~\ref{table}).  For both
Models~2 and 3, we find that the spin parameter is moderate ($<0.75$).
Meanwhile, the disc ionisation has pegged at its maximum value
($\xi=10^4\ergpcmsqps$) indicating that the surface layer of the
accretion disc is highly ionised, with iron possibly being fully
ionised.  In such circumstances, the Fe absorption edge can be
particularly strong and is often found to be highly smeared (see
\citealt{rossfabianbrandt96} and references therein).

\begin{table*}
\caption{\asca 1-10~\kev\ spectral fit parameters with a variety of reflection-based models.}
\label{table}
\begin{tabular}{lcccccccc}                
  \hline
  \hline 
& Model~1 & Model~2 & Model~3 & Model~4 & Model~5\\
& & & & &\\

%Parameters & \simplb$\otimes$\diskbb+{\sc laor} & \simplr$\otimes$\diskbb+\kerrconv$\otimes$\reflionx & \simplr$\otimes$\kerrbb+\kerrconv$\otimes$\reflionx & \powerlaw+\kerrconv$\otimes$\refbhb \\ 
Parameters & \simplb$\otimes$\diskbb & \simplr$\otimes$\diskbb & \simplr$\otimes$\kerrbb  &    \multicolumn{2}{c}{\powerlaw} \\ 
    & +{\sc laor} &       +\kerrconv$\otimes$\reflionx & +\kerrconv$\otimes$\reflionx &  \multicolumn{2}{c}{+\kerrconv$\otimes$\refbhb} \\ 
  \hline

\nnh\ ($\times10^{22}$\pcmsq) & $0.576^{+0.003}_{-0.002}$ & $0.650^{+0.006}_{-0.002}$ & $0.666^{+0.002}_{-0.007}$ & $0.663^{+0.002}_{-0.006}$ & $0.653^{+0.009}_{-0.007}$\\
$\Gamma$  & $2.40\pm0.01$ & $2.329^{+0.006}_{-0.010}$& $2.320^{+0.003}_{-0.002}$ &$2.24\pm0.01$&$2.22^{+0.03}_{-0.02}$\\
$f_{\rm SC}$ ($N_{hard}$)$^a$ & $0.6\pm0.3$ & $0.64\pm0.04$& $0.616\pm0.002$ & $2.4\pm0.1$& $2.3^{+0.2}_{-0.1}$\\
$kT$ (\kev)  & $0.513^{+0.009}_{-0.006}$ & $0.566^{+0.001}_{-0.013}$ &---& $0.540\pm0.001$& $0.542^{+0.002}_{-0.001}$ \\
$N_{diskbb}$ ({\tiny$(\tfrac{R/km}{D/10~kpc})^2cos~i $})& $5200^{+500}_{-200}$ & $4211^{+86}_{-272}$&---&---&---\\
$\Mdot$ ($\times10^{18}~{\rm g~s^{-1}}$) &--- &---&
$0.668^{+0.003}_{-0.03}$&---&---\\
$q$ &$2.0\pm0.2$ & $1.88\pm0.01$  & $1.85^{+0.2}_{-0.30}$ & $2.38^{+0.04}_{-0.07} $& $2.5^{+0.2}_{-0.1} $\\
$i$ (degrees)  & 71--78& 71--78& 71--78 & $77\pm1$& $82_{-3}$\\
$E_{Laor}$ (\kev) & $6.40^{+0.01}$ & ---& --- &---&--- \\
\rrin\ (\rrg) &$8.2^{+2.9}_{-3.5} $ & ---&---&--- &---\\
$N_{laor}$ ($\times10^{-3}$) & $7.1\pm0.1$ &--- &---&---&---\\
$\xi$ ($\ergcmps$) &  --- & $ 10000_{-320}$ & $ 10000_{-900}$& ---&--- \\
$N_{reflionx}$ ($\times10^{-6}$) & ---&$1.32\pm0.06$ &$1.288^{+0.004}_{-0.070}$& ---&---\\
$H_{den}$ ($\times10^{22} {\rm H~cm^{-3}}$) & ---& --- &---& $1.00_{-0.02}$& $1.00_{-0.02}$\\
$F_{illum}/F_{bb}$ & --- & --- & ---&$0.29^{+0.03}_{-0.18}$&$0.25^{+0.08}_{-0.07}$\\
$N_{refbhb}$ ($\times10^{-2}$) & ---&---& ---& $6.2^{+0.2}_{-2.7}$& $5.96^{+0.5}_{-1.1}$\\
spin ($\spin$) & --- & $<0.75$ & $0.45(<0.75)$ & $0.6(>0.38)$& $0.55^{+0.15}_{-0.22}$\\
$\chi^{2}/\nu$& 1848.0/1501 & 1752.3/1499 & 1759.5/1499 & 1700.9/1498 &1698.6/1498\\
  \hline 
  \end{tabular}
\begin{flushleft}
\small{ Notes: All errors are quoted at the 90 per~cent confidence level for one parameter of interest $\Delta\chisq=2.71$). Model~1, which is purely phenomenological, uses the familiar \laor\ line and allows a comparison with previous work.  Models~2 and 3 use different disc components; however both of them employ the same full reflection model (\reflionx), while treating the Compton component using \simplr\ (\S~\ref{subsec:simplr}). The core of Model~4 is \refbhb\, which is likewise a full reflection model, with the added virtue that it self-consistently models the thermal component as well.  In Models~1 to 4 the inclination was constrained to be between 71 and 78 degrees. In Model~5 the inclination is allowed to range from 60 to 82 degrees.  A constraint on the inclination was achieved only for Models 4 and 5.  \\ $^a$ The {\sc powerlaw} normalisation is in photons cm$^{-2}$ s$^{-1}$ for Models 4 and 5.  For Models 1--3, the normalisation is given by the dimensionless parameter $\fsc$ (see \S~\ref{section:analysis}). }
\end{flushleft}
\end{table*}

In order to incorporate the effects of thermal ionisation expected for
a hot accretion disc, we replace \reflionx\ with the model \refbhb\
developed by \cite{refbhb}.  This reflection model accounts for both
thermal X-ray emission and the reflection features.  The effects of
Compton broadening in the disc are fully included, subject to the one
assumption of a constant-density atmosphere.  The parameters of the
model are the number density of hydrogen in the illuminated surface
layer, ${\it H}_{\rm den}$, the temperature of the blackbody heating
the surface layers, the power-law photon index, and the ratio of the
total flux illuminating the disc to the total blackbody flux emitted
by the disc. Again, we tie the power-law index of \refbhb\ to that of
the Compton component -- now modelled as a standard power law -- and
convolve the spectrum with \kerrconv\ in order to include relativistic
broadening. The model results in an excellent fit to the data with
$\chi^{2}/\nu = 1700.9/1498$ (Model~4, see bottom panel in
Fig.~\ref{fig_ratio_ref}), however the hydrogen surface density is
pegged at the maximum value of the model.  The ionisation state of the
disc is inversely proportional to the value of the hydrogen density
and thus the pegged value implies that the fit is requiring a higher
amount of emission in the form of discrete features as opposed to the
near featureless reflected continuum arising from a highly ionised
disc-surface. A similar result would be produced by increasing the
iron abundance. Unfortunately the model in its current format does not
allow for a change in elemental abundances. In order to investigate
the effect that $H_{den}$ has on the spin parameter we fixed it at
$1\times10^{21}~{\rm H~cm^{-3}}$ using Model~4 (i.e. an order of
magnitude less than the value presented in Table~\ref{table}) and
refitted the data. This constraint on $H_{den}$ resulted in an
adequate fit with $\chi^{2}/\nu = 1731.6/1499$ and a spin value of
$0.60\pm0.05$.  We note here that {\sc powerlaw} has been used to
model the Compton component.  We have explored replacing {\sc
  powerlaw} with \simplr, and the fit becomes worse with $\chi^{2}/\nu
= 1802.5/1498$. However, the value of the spin parameter, as well as
those of the reflection parameters, remains largely unchanged.

\begin{figure*}
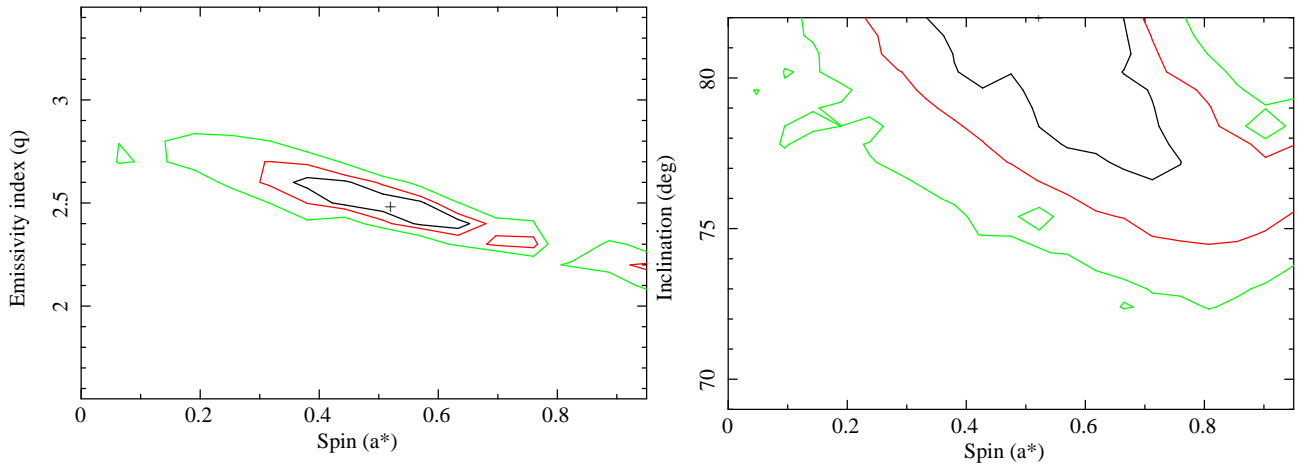

\centering{\includegraphics[angle=270, width=8.5cm]{fig10a.eps}}
\vspace{0.5cm}
\centering{\includegraphics[angle=270, width=8.5cm]{fig10b.eps}}
\caption{ ({\it left}): Emissivity versus spin contour plot for \j. The
  68, 90 and 95 per cent confidence range for two parameters of
  interest are shown in black, red and green, respectively.  We have
  allowed $i$ to take any value between $60\leq i \leq82$, and find
  that the spin is greater than $0.33$ at the 90~per cent level of
  confidence. ({\it right:}) Similar plot for inclination versus
  spin. We see from the \asca data that a zero spin value is clearly
  ruled out, as is an inclination lower than $72\degr$. }
\label{fig_contour}
\end{figure*}

\begin{figure}
\centering {\includegraphics[angle=0, width=8.8cm]{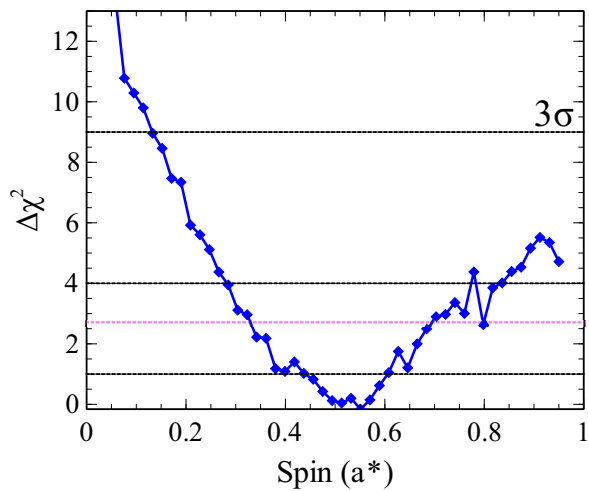}}
%{figure_steppar_spin_1550.eps}}
\caption{ Goodness-of-fit versus spin parameter for \j. From the
  reflection features present in the \asca spectra of \j\ we can rule
  out a non rotating black hole at over 3$\sigma$ confidence.  However
  we cannot place a comparable strong upper limit on this value.  The
  90~per cent confidence level ($\Delta\chi^2=2.71$ for one parameter
  of interest) is shown in magenta. The black dotted lines indicate
  confidence intervals. Spin is constrained to $0.33 < \spin < 0.70$
  at the 90 per~cent confidence level. }
\label{fig_steppar_spin1}
\end{figure}

From Models~1 to 4 it is clear that the spin parameter is consistently
below $\approx0.75$. However, in the first three cases inclination is
not constrained.  For this reason we explore a very broad range of
$i$, from 60\degr\ to 82\degr.  We note that above this limit, the
disc would be super-Eddington during its steady thermal plateau in
Figure~\ref{fig:state-lc}.  The best fit is given by Model~5
(Tab.~\ref{table}) and reaches the upper inclination limit, netting a
small improvement ($\Delta\chisq = -2.3$) over Model~4.  For all
models we see that the emissivity index is consistently below the
typical value of three associated with the canonical `lamp-post'
coronal geometry, and is instead more consistent with a slab-like
corona.  In order to illuminate any degeneracy between the value of
spin and either the emissivity index or inclination, we show in
Figure~\ref{fig_contour} the 68, 90 and 95~per cent probability
contours for these parameters plotted versus spin.  In both instances
there exists a small and negative correlation with spin.  However it
is also clear that $q$ is well constrained between 2.2 and 2.7 and
that $i \gtrsim 75\degr$ at 90 per~cent confidence even while including the
uncertainty in spin.  When we marginalize over these parameters and
compute the uncertainty in spin alone (Figure~\ref{fig_steppar_spin1}
for Model~5), the spin parameter obtained from the gravitational
blurring of reflection features is constrained to be in the window
$0.33 < \spin < 0.70$ at 90 per cent confidence with the best estimate
at $\spin \approx 0.55$.  A non-rotating Schwarzschild black hole is
rejected at greater than 3$\sigma$.

Our measured spin using the \refbhb\ model is consistently lower than
the preliminary value of $\spin = 0.75-0.80$ reported by
\citet{Miller_2009}, and we have attained a better fit than they
($\Delta\chisq \lesssim -100$) for more degrees of freedom.  The
critical difference in our model and spin estimate comes from having
incorporated the effect Compton-broadening of the iron~\ka\ line in
the hot layers of the accretion disc. With \refbhb\ the disc is
intrinsically hot and therefore the effect of Compton-broadening is
fully accounted for when modeling the data. The extra broadening
caused by this effect acts to lower the degree of gravitational
broadening and as such requires less extreme spin parameters as
compared with models where the reflection is assumed to come from a
relatively cold surface, e.g, for AGN \citep{refbhb}. This behavior
can indeed be appreciated when we compare the spin value obtained from
\reflionx\ -- a reflection model specifically designed for a cold
accretion disc -- to that of \refbhb. We see from Table~\ref{table}
that for Models~2 and 3 the spin cover a higher range, with the
90~per cent error going as high as 0.75.

\subsection{Spin from reflection features -- \rxte}
\label{ref_rxte}

\begin{figure}
\centering{\includegraphics[width=7.5cm]{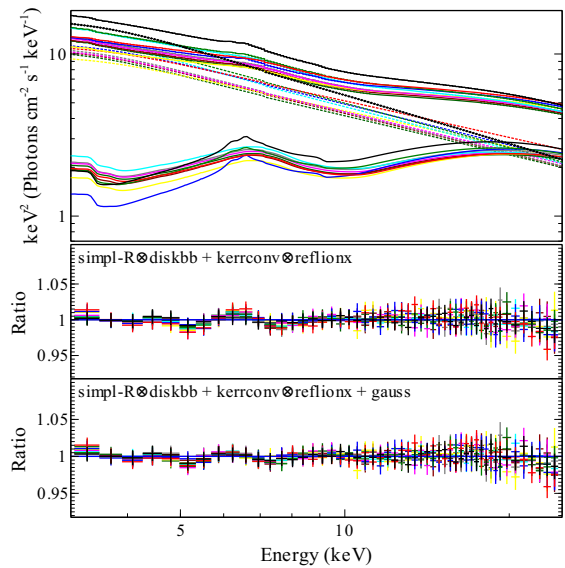}}
\caption{Best-fitting model ({\it top}) for the ten \rxte
  observations. The model consists of a disc and Compton continuum
  together with a relativistically blurred reflection component. The
  spin, inclination, and emissivity index were treated as global
  parameters (see \S~\ref{ref_rxte}). ({\it middle:}) Data/model ratio
  for the above model and ({\it bottom:}) after the inclusion of a
  narrow Gaussian emission line. }
\label{fig_rxte_model}
\end{figure}

In order to supplement the \asca spin measurement above, we present an
analysis of a sample of ten \rxte spectra selected from the composite
data set discussed in
\S\S~\ref{section:analysis},\ref{section:self-consistent} to have the
following properties: very large scattered fraction, $\fsc > 50$ per~cent,
goodness of fit, $\rchinu < 2$, and uniform values of luminosity and
photon index, $\lledd \approx 0.2 \pm 0.05$ and $\Gamma \approx 2.5
\pm 0.1$, respectively.  We begin by simultaneously modeling the
reflection features present in all the \rxte spectra using \reflionx\
convolved with \kerrconv\ while using \simplr$\otimes$\diskbb\ for the
thermal plus Compton continuum (Model~2 in \S~\ref{full_refl}).
(N.B. The \refbhb\ component used in Models~4 and 5 was unable to
converge to an adequate fit for \rxte and provided no spin constraint.
Therefore, in this section we adopt Model~2.)  The spin, inclination
($60\leq i \leq82$) and emissivity index{\footnote{Because we have
    selected a homogeneous set of spectra with almost identical
    luminosities, it is likely that the emissivity index -- an
    indicator of coronal geometry -- is the same for all ten
    spectra.}} are treated as global parameters among the ten
spectra. As in \S~\ref{section:analysis}, the neutral hydrogen column
density is fixed at $8\times10^{21}\pcmsq$.  The remaining parameters
were allowed to vary in individual spectra.
Figure~\ref{fig_rxte_model} shows the best-fitting model spectra (top
panel) together with the data-to-model ratio for each spectrum (bottom
panel).  The fit was marginally improved ($\Delta\chi^2=-30.5$ for 10
degrees of freedom) by including a narrow line at $\approx$6.7~\kev\,
which accounts for the slight curvature in the residuals at that
energy (compare the lower two panels in Fig.~\ref{fig_rxte_model}).

We find that the global best fit is sensitive to the upper energy
range adopted for the \rxte spectra, which we attribute in part to a
competition between the lower energy reflection features and the
high-energy Compton hump.  Considering upper ranges between
12--45~\keV, the best spin estimate was found between $\spin \approx
0.6-0.69$ giving reduced chi-square values from $\rchinu = 0.4-0.8$,
with higher values obtained at extended energy ranges.  Most
importantly, the model consistently estimated the 90 per~cent upper limit for
spin at $\spin = 0.75$.  For the other global parameters, we treat the
\rxte\ results as second-tier, but find results consistent with the
\asca values: $q \approx 2.5$ and $i > 72\degr$ (90 per~cent).

We are cautious in interpreting this spin estimate using \rxte
spectral fits, owing to the coarse ($\sim 20$ per~cent) energy resolution.
However, we expect that \rxte should provide robust upper bounds on
the degree of relativistic broadening (viz., spin), owing to its vast
collecting area and $\lesssim 1$ per~cent spectral calibration
\citep{Jahoda_2006}.  We caution towards the significance of the
\rxte-derived spin parameter and consider the upper limit obtained
here as a {\em complementary result} to that obtained from the \asca
data alone, confirming that Fe-\ka\ spin is not high.

%\clearpage}

%%%%%%%%%%%%%%%%%%%%%%%%%%%%%%%%%%%%%%%%%%%%%%%%%%%%%%%%%%
%%%%%%%%%%%%%                     %%%%%%%%%%%%%%%%%%%%%%%%
%%%%%%%%%%        END    RUBENS      %%%%%%%%%%%%%%%%%%%%%
%%%%%%%%%%%%%                     %%%%%%%%%%%%%%%%%%%%%%%%
%%%%%%%%%%%%%%%%%%%%%%%%%%%%%%%%%%%%%%%%%%%%%%%%%%%%%%%%%%

\section{Discussion}\label{section:discussion}

\subsection{A Combined Fe-\ka\ and CF Result}\label{subsec:comparison}

In the two previous sections, we concluded that both the Fe-line and
CF methods predict moderate values of spin, which are quite
consistent: $0.33 < \spin < 0.70$ (\asca only) and $-0.11 < \spin <
0.71 $, respectively (90 per~cent confidence).  The CF spin result predicts a
slightly narrower Fe-line feature than that found by the Fe-line
analysis.  Alternatively, the Fe-line measurements consistently favor
a high inclination, and therefore require a lesser distance
($D\approx$4~kpc), in order for the CF results to match.

\begin{figure}
\centering{\includegraphics[clip=true, angle=90,width=9.5cm]{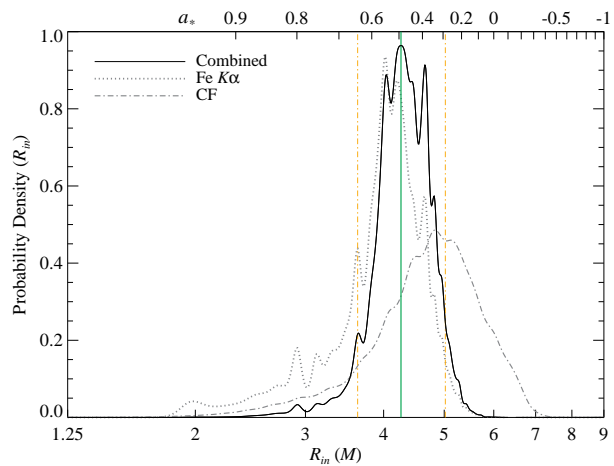}}
\caption{Combined Fe-\ka\ and CF probability density for $\Rin$ and
  $\spin$.  The net result is again a moderate value in-between the
  two individual estimates ($\spin = 0.49^{+0.13}_{-0.20}$, 90 per~cent
  confidence).}\label{fig:combospin}
\end{figure}

Having obtained two independent measurements of the spin, we now
combine them by convolving the individual spin probability
distributions to obtain the joint distribution shown in
Figure~\ref{fig:combospin}.  Our synthesized result is then $ 0.29 <
\spin < 0.62$, with a most probable value of $\spin = 0.49$.
Remarkably, based on a model of binary evolution and the GRB collapsar
model, \citet{Brown_2007} predicted that \j\ formed with $\spin
\approx 0.5$.  We confirm their prediction.

\subsection{Confronting GRMHD Simulations}\label{subsec:insideisco}

Recently, it has become feasible, via general-relativistic
magnetohydrodynamic (GRMHD) simulations, to assess the differences
between MRI-driven accretion flows and the idealised $\alpha$-disc
model, upon which our CF model is based.  These differences include
both a non-zero torque inside the ISCO and an altered angular momentum
profile for the disc.  The results of \citet{Penna_2010} show, for the
geometrically thin accretion discs we consider, that these effects are
expected to impact our results by $\lesssim 7$ per~cent in $\rin$ (roughly
$\Delta\spin \lesssim 0.09$ here).  In general, they conclude that deviations
from the Novikov-Thorne model are small for thin discs at low
luminosity and grow worse as the luminosity rises and the disc
thickens.

Similar MHD simulations have been made to assess a principal
assumption of the Fe-line method, namely, that the line emission from
within the ISCO is negligible \citep{Reynolds_Fabian_2008}.  Including
the effect of contributing plunging-region emission results in
intrinsically broader line profiles and hence will lower the estimate
for spin.  For the disc thickness and spin values in question,
simulations predict that this effect could possibly shift $\rin$ by
$\sim 12$ per~cent (value taken from Fig.~5 in
\citealt{Reynolds_Fabian_2008}), thereby decreasing the most probable
Fe-\ka\ estimate of spin from $\spin \approx 0.55$ to $\approx 0.4$,
in very close agreement with the best CF value of $\spin \approx
0.34$.

We infer that both spin estimates are subject to moderate corrections.
For the Fe-\ka\ reflection method, we can specifically conclude that the
corrected value of spin is lower than the measured value, strengthening
our conclusion that the spin of \j\ is moderate.

\subsection{The Question of Alignment}\label{subsec:alignment}

The spin of an accreting black hole in a binary is expected to align
with the orbital angular momentum vector of the system within $\approx
10^7 - 10^8$ years \citep{Maccarone_2002}.  A recent population
synthesis study, which makes conservative assumptions concerning the
torques acting to align a black hole, predicts that most black holes
will be aligned to better than 10$\degr$.  In \S~\ref{reflection}, we
constrained the inclination of the inner, reflecting portion of the
accretion disc (Fig.~\ref{fig_contour}).  This allows us to check on
the relative alignment of the black hole spin axis (which is aligned
with this inner-disc region; \citealt{Lodato_Pringle_2006}) and the
orbital vector.  In our exploration of the Fe-\ka\ model, for a wide
range of orbital inclinations (60\degr -- 82\degr), we find a best-fitting
inclination for the inner disc of $\approx75-82$\degr.  This value is
consistent with the orbital inclination angle given by our dynamical
model, $i=74.7\pm3.8$ \citep{Orosz_Steiner_2010}, which validates the
CF assumption of alignment \citep{KERRBB}, while simultaneously
providing support for the dynamical model.

%  Considering the high orbital inclination
%  predicted for \j, we find it unlikely that a large misalignment
%  could persist without causing prominent X-ray variability caused by
%  the Bardeen-Peterson misalignment between inner-and-outer accretion
%  discs.  Such an effect has not been seen in \j, and so we consider
%  this further indication that a sizable misalignment is unlikely to
%  be present.} \\
%

\subsection{Implications of a Low-Spin Microquasar}

The low spins of J1550 and the microquasar A0620--00 ($a_*\approx0.1$;
\citealt{Gou_2010}) challenge the long-standing and widely-held belief
that there is a strong connection between black hole spin and
relativistic jets (\citealt{BZ77}; hereafter BZ).  We note that this
belief is also challenged by a statistical study that found no
evidence for a link between black hole spin and jet power
\citep{Fender_2010}.  In any case, if jets are powered by black hole
spin, then theory predicts that jet power will increase dramatically
with increasing $a_*$ \citep{Tchekhovskoy_2010}, with the jet
receiving more power from the accretion disc than from the black hole
for $a_*>0.4$ \citep{McKinney_2005}.

Given the low spins of J1550 and A0620--00, it would appear that their
episodic jets are driven largely by the accretion disc.  One
well-known candidate mechanism is the centrifugally driven outflow of
matter from a disc described by \citealt{BP82} (hereafter BP).  A
useful comparison of the operational regimes of BP and BZ is given by
\citet{Garofalo_2010}.  They show that BP is always viable, but that
BZ is a more likely mechanism for the most rapidly rotating sources,
such as the extreme-Kerr BH microquasar GRS~1915+105
\citep{McClintock_2006, Blum_2009}.

\begin{figure}
\centering{\includegraphics[clip=true, angle=90,width=9.8cm]{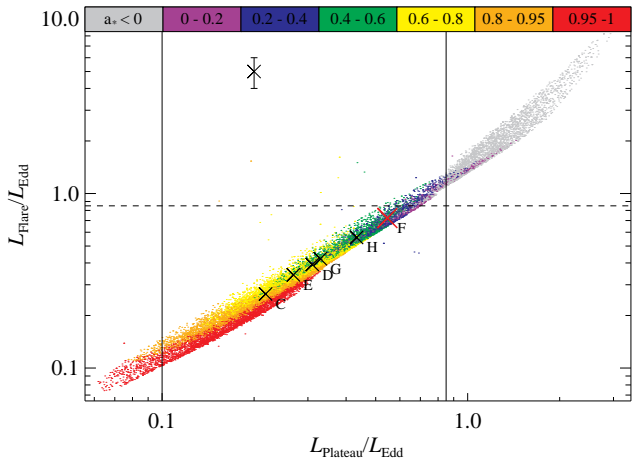}}
\caption{The intrinsic (i.e., seed) luminosity of the disc component
  during the 7 Crab flare versus the luminosity during the thermal
  plateau phase.  In order to avoid saturating the plot, we show only
  half the data points, which were selected at random.  The vertical
  black lines mark the lower and upper luminosity thresholds, and the
  horizontal dashed line corresponds to the Eddington limit of an
  accretion disc (\S~\ref{section:systematics}).  Note that Model F is
  very near this limit. }
\label{fig:luminositybanana}
\end{figure}

The relativistic, two-sided jet of J1550 was launched during the
remarkable 7 Crab flare (see \S~\ref{section:Intro}).  We now show
that during this daylong event the luminosity of the accretion disc
was close to, or perhaps at, its Eddington limit, which suggests that
radiation pressure was a key agent in collimating or feeding a
magnetically-accelerated jet via a disc wind (for a discussion of the
interplay between jet and wind, see \citealt{Neilsen_naturally} and
\citealt{Miller_2008}).  In Figure~\ref{fig:luminositybanana}, we plot
the intrinsic accretion-disc luminosity during the flare state versus
the luminosity during the thermal-dominant plateau state (Days
105--182; see Fig.~\ref{fig:state-lc}).  Each data point represents a
single triplet of values of $M$, $i$ and $D$ from among the 42,500
triplets considered in our Odyssey cluster analysis (Appendix A.4),
and that data point was derived by analysing the complete data set for
J1550.  The spin for each point (averaged over the \gold\ and \silver\
data) is indicated by a colour, which is encoded in the bar at the top
of the plot. The point corresponding to the dynamical model adopted by
\citet{Orosz_Steiner_2010}, Model F, which is our fiducial model, is
labelled and marked by a red cross.  The five less probable models
considered by Orosz et al.\ are marked by black crosses.  We conclude
that Model F, by far the most probable model (see Appendix A.5), is
very near the Eddington limit for disc accretion.

Luminous discs are geometrically thick and widely believed to be
effective at driving jets.  In the case of the \j\ flare, the disc is
not only thick, it is also near its Eddington limit, so that it will
provide substantial radiation pressure and possibly even shed material
via a radiation-driven outflow, thereby promoting a jet-ejection
event.  In any case, as a bottom line, the low spins of \j\ and
A0620--00 indicate that spin is not the sole driver behind all
powerful episodic jets.

\section{Conclusion}\label{section:conclusion}

For the first time in a single work, we have determined high-quality
estimates of the spin of an accreting black hole using the two
independent, leading methods.  In our CF analysis, we carefully
explored the sensitivity of our results to a wide range of
model-dependent systematic errors and observational errors.  We
conclude that \j\ is a slowly spinning black hole with $\spin \approx
0.34$, while ruling out spins larger than $\spin \gtrsim 0.71$ at 90 per~cent
confidence.  Next, we analysed the Fe-\ka\ and reflection signatures
in bright, intermediate spectral states of \j.  By modeling these
broad, skewed features, we obtained a slightly higher estimate of the
spin, $\spin \approx 0.55^{+0.15}_{-0.22}$ (at 90 per~cent confidence), while
also deriving an estimate of the inclination angle of the inner disc
that is in close agreement with the orbital inclination angle
\citep{Orosz_Steiner_2010}.  Combining the two spin estimates, we
conclude that J1550, like the microquasar A0620--00, is a slowly
spinning black hole.

The low spins of both J1550 and A0620--00 indicate that, for at least
some microquasars, BZ-type mechanisms are not primary in driving
powerful episodic jets, and that other mechanisms (perhaps BP) are at
play.  The near Eddington-limited 7 Crab flare observed for J1550
suggests that radiation-pressure support from a thermal disc is one
possible way that low-spin black holes are aided in driving large-scale
relativistic jets.

\section*{Acknowledgments}

JFS was supported by the Smithsonian Institution Endowment Funds and
JEM acknowledges support from NASA grant NNX08AJ55G.  RN acknowledges
support from NASA grant NNX08AH32G and NSF grant AST-0805832.  RCR
would like to thank the Science and Technology Research Council (STFC)
and the Smithsonian Astrophysical Observatory for financial support.
JFS thanks Bob Penna and Sasha Tchekhovskoy for helpful discussions,
and the FAS Sciences Division Research Computing Group for their
technical support with analyses performed on the Odyssey cluster.

\appendix

\section{Continuum-Fitting: Assessing the Systematic Uncertainties}\label{append:systematics}

\subsection{Model Parameters}\label{subsec:systematics:params}

We consider the effect of the principal parameters listed in
Table~\ref{tab:systematics} on our final determination of the spin
(Figure~\ref{fig:spin_mid_gauss}) for Model~S
(\S~\ref{section:analysis}) and Model~I
(\S~\ref{section:self-consistent}).  Here and below, we consider only
the \gold\ data.  As in
\S\S~\ref{subsec:resultsI},\ref{subsec:ireflect}, we fix $M$, $i$, and
$D$ at their fiducial values (\S~\ref{section:Intro}).  As illustrated
in Figure~\ref{fig:spin_mid_gauss}, the effect of decreasing the
viscosity parameter (P1 in Table~\ref{tab:systematics}) is to decrease
$\rin$ (by $\sim 3-6$ per~cent, depending on the model).

We allow the column density (P2--3) to vary over a broad range, $\nh =
6-10 \times 10^{21}~\pcmsq$, which corresponds to $\gtrsim 8\sigma$
relative to the precise value determined using {\it Chandra} grating
data (see \S~\ref{subsec:resultsI}).  We consider this extreme range
because of the discrepant results for $\nh$ obtained using \asca data
(see \S~\ref{reflection}), which we attribute to an error in the
calibration of the \asca detectors at low energies.  As shown in
Table~\ref{tab:systematics}, our liberal estimate of the uncertainty
in $\nh$ affects our determination of $\rin$ by $<3$ per~cent.

We next explore the parameters of the \ireflect\ model.  We test
smaller covering factors of 1/2 and 1/10 by linking the covering
factor in \ireflect\ to $-(x-1)$ from the \simplr\ model, where the
leading minus sign acts as a switch in the model to isolate reflection
from the direct (illuminating) component.  Thus, we consider two
cases: $x=1.5$ (P4) and $x=1.1$ (P5).  We next try fitting for the
covering factor, allowing it to vary between 0 and 1, while also
fitting for the emissivity, which we constrain to lie in the range
$2<q<5$ (P6).  As shown in Table~\ref{tab:systematics}, the effect of
this exercise on $\rin$ is small $\approx1$ per~cent.  Smaller still is the
effect of varying the disc temperature.  Decreasing $T_{\rm disc}$
(P7) by a factor of 5 relative to its assumed value
(\S~\ref{subsec:ireflect}), we find that the ionisation parameter
increases slightly, but that the effect on $\rin$ is negligible
($<0.1$ per~cent).

Lastly, we adjust the width of the \smedge\ component $W_{\rm Edge}$
to first half (P8) and then twice (P9) its nominal value of 7 keV,
which impacts $\rin$ by $<0.5$ per~cent.

In summary, as we found earlier in our study of LMC X-3
\citep{Steiner_lmcx3}, $\alpha$ is the parameter (aside from $M$, $i$,
and $D$) that introduces the largest uncertainty in determining spin
via the CF method.

\subsection{Model Components}\label{subsec:systematics:model}

We begin by substituting \bhspec \citep{BHSPEC} for the thermal disc
component in place of \kerrbbtwo (see \S~\ref{subsec:ireflect} in
\citealt{McClintock_2006} for a discussion of these relativistic disc
models and their relationship).  The virtue of \bhspec relative to
\kerrbbtwo is that it directly computes the effects of spectral
hardening; its drawback is that it does not include returning
radiation, which heats the disc.  Employing \bhspec and following the
procedures described in the preceding section, we find that $\rin$ is
increased by $\approx 1-3$ per~cent (M1 in Table~\ref{tab:systematics}).

Next, we explore the possibility that the power-law component is cut
off exponentially at high energy (e.g., thermal Comptonisation), while
allowing the cut-off energy to vary over the range $kT_e=25-200~\keV$.
(We did not correct $\fsc$ in order to achieve photon conservation
because this correction is negligible for the \gold\ spectra.)  We
find that the effect of a possible cutoff is small, changing $\rin$ by
$\lesssim 1.5$ per~cent (M2).  In addition, we generated the power law using
the double-sided version of \simpl and \simplr\ (in place of the
upscattering-only version).  The effect on $\rin$ is $\lesssim 0.3$ per~cent
(M3).

Lastly, we examine the effects of substituting one of the reflection
components for another.  We find that both \ireflect\ and \reflionx\
give somewhat smaller values of $\rin$ than \smedge, but the effect is
small, $\lesssim 3$ per~cent (M4, M5).

\subsection{Flux}\label{subsec:systematics:flux}

As described in \S~\ref{section:systematics}, we include a liberal
$\sim10$ per~cent allowance for the uncertainty in the absolute-flux
calibration.  Because the luminosity of the thermal component at a
given colour temperature scales proportionally to $\rin^2$, a 10 per~cent
adjustment to the flux normalisation introduces a 5 per~cent uncertainty in
$\rin$.

\subsection{Black Hole Mass, Inclination and Distance}\label{subsec:systematics:mid}

While analysing the X-ray spectral data, we have used the best
estimates for $M$, $i$ and $D$ (\S~\ref{section:Intro}) taken from to
Model F of Table~1 in \citet{Orosz_Steiner_2010}.  To explore the
sensitivity of spin to uncertainties in these measured values, we use
the Odyssey computing cluster at Harvard University and fit the data
set at each point in a 3-D grid of 42,500 points distributed uniformly
over mass, inclination, and distance.  The grid spans the ranges
$M=5-17.5~\msun$, $i=36-85^\circ$ and $D=3-7~\kpc$, respectively.  We
adopt the 3~kpc distance bound following \citet{Hannikainen_2009}; the
7~kpc bound is a relativistic limit based on the proper motion of the
X-ray jets \citep{Corbel_2002}: $D \leq c/\sqrt{\mu_a \mu_r} \lesssim
7~\kpc$ (e.g., \citealt{Mirabel_Rodriguez}).  At each grid point, we
compute a table of the spectral hardening factor (e.g., see
\citealt{Gou_2010}) and fit all of the available TD, SPL, and INT
spectra.  We use Model~S (\S~\ref{subsec:resultsI}) because it is
computationally efficient and has the best performance of all three CF
models considered.  We perform the analysis for both values of disc
viscosity: $\alpha = 0.01$ and $\alpha = 0.1$.

We then apply our data selection criteria, obtaining a sample of
\gold\ and \silver\ spectra (typically 50--100) at each of the 42,500
grid points.  From this we derive a spin probability distribution
unique to each point.  Before summing over the grid, we impose the
following grid point selection constraints: First, the grid point's
inclination must be below the eclipsing limit, $i < 82\degr$ (see
e.g., \citealt{Narayan_McClintock_2005}).  Secondly, as discussed in
\S~\ref{section:systematics}, we require that the intrinsic disc
luminosity during the TD-state plateau phase (days 105--181;
Figure~\ref{fig:state-lc}) fall in the range $0.10 < L_{\rm D}/L_{\rm
  Edd} < 0.85$.  We combine the distributions for all satisfactory
grid points, weighting each according to its location in the grid
(with high weights occurring at probable values of $M$, $i$, and $D$).

\subsection{Rolling Together the Uncertainties}

We combine the systematic uncertainties discussed above in two stages.
Referring to Table~\ref{tab:systematics}, in the first stage we
combine in quadrature the individual values in the Model~S column for
rows P8-9, M1, M2-3, M4-5 with half the value for P2-3 (half because
the range of variation considered for $\nh$ is so extreme).  For each
of the parameters $\nh$ and $W_{\rm Edge}$, we use the larger of the
deviations given in the table.  The resultant error of 4.2 per~cent is
combined with the 5 per~cent error in $\rin$ from flux uncertainty to give a
net error of 6.5 per~cent.  This combined uncertainty sets the half-length
for a boxcar smoothing kernel that we apply to the full spin
distribution.

\begin{figure}
\centering{\includegraphics[clip=true, angle=90,width=8.8cm]{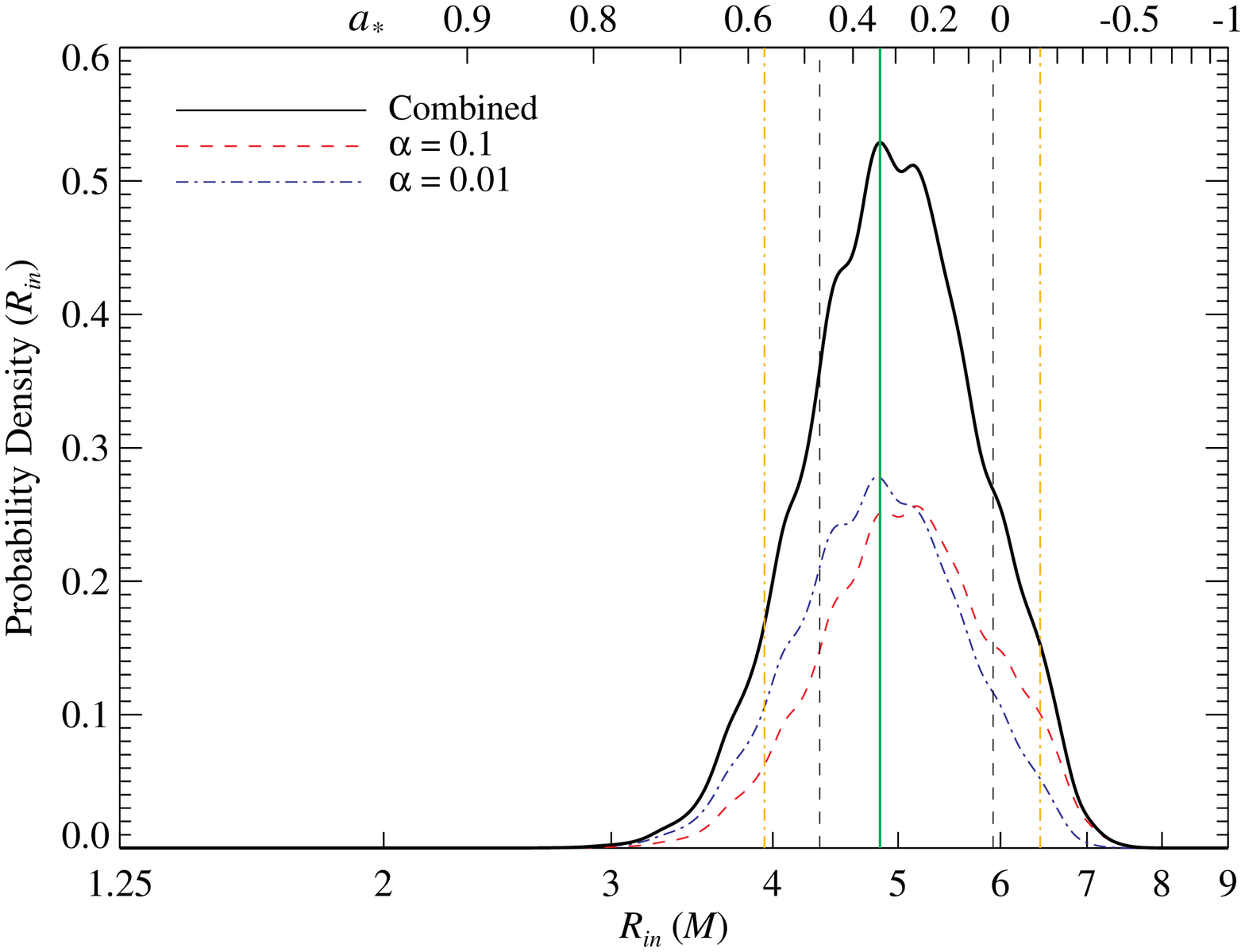}}
\caption{Similar to Figure~\ref{fig:spin_mid_gauss}, but using just
Model~F in \citet{Orosz_Steiner_2010}.  The green, black, and gold
vertical lines indicate the most likely value for $\Rin$ ($\spin$),
and $1\sigma$ and 90 per~cent confidence interval limits,
respectively.}\label{fig:bestmod}
\end{figure}

The resulting distribution is shown in Figure~\ref{fig:bestmod}.
Because we have so far considered just dynamical Model~F, the
distribution of $\rin$ is narrower, $-0.14 < \spin < 0.57$ (90 per~cent
confidence level) than our final distribution shown in
Figure~\ref{fig:spin_mid_gauss}, although the most probable value of
spin is unchanged, $\spin = 0.34$.  We now go on to the second stage
in combining sources of error and consider an ensemble of possible
dynamical models.

The case of \j\ is unusual in that there are several candidate models
which produce reasonable fits to the dynamical data, which are
summarized in Table~1 in \citet{Orosz_Steiner_2010}).  Above, we
considered only Model~F, the most probable model.  We now incorporate
the possibility that one of the five alternative models (Models C-E
and Models G \& H) are correct.  Models A and B do not constrain the
dynamical model satisfactorily, and do not allow one to obtain a
useful distance estimate, and so they are disregarded here.  

As was done above for Model~F, a spin ($\rin$) probability
distribution is obtained for each candidate dynamical model.  We use
for each model, including Model~F, the total $\chisq$ (summed for the
velocity data and the light curve data, both optical and infrared) to
determine its corrected Akaike Information Criterion (AICc;
\citealt{AIC, AICc}), which is closely related to the log-likelihood
of each model.  Using these values, AIC-weights are assigned to each
model ($i$): $W_{{\rm AIC}, ~ i} = {\rm Exp[-1/2~ (AICc}_i - {\it
    inf}\{{\rm AICc }\})]$ \citep{Burnham_Anderson}.

Our fiducial dynamical model is by far the most likely, carrying $\sim
84$ per~cent of the total weight.  A weighted sum is computed using the
AIC-weights to obtain a composite spin distribution.  This is
broadened using the boxcar smoothing kernel described above (13 per~cent
width) to produce the final distribution as shown in
Figure~\ref{fig:spin_mid_gauss}.  {\em Thus, this final result
  incorporates uncertainties in the choice of the dynamical model;
  dynamical model uncertainties; the X-ray spectral model and model
  parameter settings as summarized in Table~\ref{tab:systematics}; and
  a 10 per~cent uncertainty in the X-ray flux calibration.}

%%%%%%%%%%%%%%%%%%%%%%%%%%%%%%%%%%%%%%%%%%%%%%%%%%%%%%%%%%%%%%%%

%%%%%%%%%%%%%%%%%%%%%%%%%%%%%%%%%%%%%%%%%%%%%%%%%%%%%%%%%%%%%%%%%%%

%\clearpage%
\newcounter{BIBcounter} % Make a counter to reference
\refstepcounter{BIBcounter}

\bibliography{js}

\end{document}